%% file: _Full_maintext.tex
\documentclass{nature2}

\usepackage{color}
\usepackage{amsmath,amssymb}
\usepackage{graphicx}
\usepackage{hyperref}
\usepackage{lineno}
\def\targname{GNz7q}

\definecolor{emph}{rgb}{0.00,0.3,0.60}
\def\tcr{\textcolor{black}}
\def\tcb{\textcolor{black}}

\newcommand{\oiii}{[O\,{\sc iii}]}

\newcommand{\cii}{[C\,{\sc ii}]}
\newcommand{\ci}{[C\,{\sc i}](2-1)}

\newcommand{\ciii}{C\,{\sc iii}]}
\newcommand{\civ}{C\,{\sc iv}}

\newcommand{\mum}{$\mu$m}

\newcommand{\lsun}{$L_{\rm \odot}$}
\newcommand{\msun}{$M_{\rm \odot}$}

\newcommand{\td}{$T_{\rm d}$}

\newcommand{\md}{$M_{\rm dust}$}

\newcommand{\mgas}{$M_{\rm gas}$}

\newcommand{\lir}{$L_{\rm IR}$}
\newcommand{\lirs}{$L_{\rm IR,SF}$}
\newcommand{\lira}{$L_{\rm IR,AGN}$}
\title{\flushleft 
A dusty compact object bridging galaxies and quasars at cosmic dawn
}

\author{\flushleft 
S.~Fujimoto$^{1,2}$,
G.~B. Brammer$^{1,2}$,
D.~Watson$^{1,2}$,
G.~E.~Magdis$^{1,3,2}$,
V.~Kokorev$^{1,2}$,
T.~R.~Greve$^{1,3}$,
S.~Toft$^{1,2}$,
F.~Walter$^{1,4,5}$,
R.~Valiante$^{6}$,
M.~Ginolfi$^{7}$,
R.~Schneider$^{6,8}$,
F.~Valentino$^{1,2}$,
L.~Colina$^{1,9}$,
M.~Vestergaard$^{2,10}$, 
R.~Marques-Chaves$^{11}$,
J.~P.~U.~Fynbo$^{1,2}$, 
M.~Krips$^{12}$,
C.~L.~Steinhardt$^{1,2}$,
I.~Cortzen$^{12}$,
F.~Rizzo$^{1,2}$, 
and P.~A.~Oesch$^{1,11}$
}

\begin{document}
\maketitle

\vspace{-0.6cm}
\tcb{
\begin{affiliations}
 \item Cosmic Dawn Center (DAWN), Copenhagen, Denmark
 \item Niels Bohr Institute, University of Copenhagen, Jagtvej 128, DK-2200 Copenhagen N, Denmark
 \item DTU-Space, Technical University of Denmark, Elektrovej 327, DK-2800 Kgs. Lyngby, Denmark
 \item Max Planck Institute for Astronomy, Konigstuhl 17, 69117 Heidelberg, Germany
 \item National Radio Astronomy Observatory, Pete V. Domenici Array Science Center, P.O. Box O, Socorro, NM 87801, USA
 \item INAF-Osservatorio Astronomico di Roma, via di Frascati 33, I-00040, Monteporzio Catone, Italy
 \item European Southern Observatory, Karl Schwarzschild Str.\ 2, D-85748 Garching, Germany
 \item Dipartimento di Fisica, Universit\'a di Roma La Sapienza P.le Aldo Moro 2, I-00185 Roma, Italy
 \item Centro de Astrobiolog\'ia (CAB, CSIC-INTA), Carretera de Ajalvir, 28850 Torrej\'on de Ardoz, Madrid, Spain
  \item Steward  Observatory, University of Arizona, 933 N Cherry  Avenue, Tucson  AZ, 85718, USA
 \item Geneva Observatory, University of Geneva, Chemin Pegasi 51, 1290 Versoix, Switzerland
 \item IRAM, Domaine Universitaire, 300 rue de la Piscine, 38406 Saint-Martin-d’H\`eres, France
\end{affiliations}
 }
\vspace{0.2cm}

\begin{abstract} 
Understanding how super-massive black holes form and grow in the early Universe has become a major challenge\cite{volonteri2012,inayoshi2020} since the discovery of luminous quasars only 700 million years after the Big Bang\cite{mortlock2011, banados2018}. 
Simulations indicate an evolutionary sequence of dust-reddened quasars emerging from heavily dust-obscured starbursts that then transition to unobscured luminous quasars by expelling gas and dust\cite{hopkins2008}. 
Although the last phase has been identified out to a redshift of 7.6,\cite{wang2021} a transitioning quasar has not been found at similar redshifts owing to their faintness at optical and near-infrared wavelengths. Here we report observations of an ultraviolet compact object, \targname, associated with a dust-enshrouded starburst at a redshift of \boldmath$ z=7.1899\pm0.0005$.
The host galaxy is
more luminous in dust emission than any other known object at this epoch, forming 1,600 solar masses of stars per year within a central
radius of 480~parsec. A red point source in the far-ultraviolet is identified in deep, high-resolution imaging and slitless spectroscopy.
\targname\ is extremely faint in X-rays, which indicates the emergence of a uniquely ultraviolet compact star-forming region or a Compton-thick super-Eddington black-hole accretion disk at the dusty starburst core.
In the latter case, the observed properties are consistent with predictions from cosmological simulations\cite{ginolfi2019} and suggest that \targname\ is an antecedent to unobscured luminous quasars at later epochs.
\end{abstract}

\clearpage

In recent uniform reprocessing of all archival \textit{Hubble Space Telescope} ({\it HST}) imaging and slitless spectroscopy (see Methods), 
\targname\ was identified in the GOODS-North extra-galactic field as a luminous galaxy candidate at a redshift $z>6.5$ with the F160W band AB magnitude of 23.09 $\pm$ 0.05.
Although the source was detected and highlighted as a potential high-redshift galaxy by previous authors, 
it was never spectroscopically confirmed\cite{hathi2012}.  
However, the full suite of {\it HST} data 
reveals an unambiguous continuum break at $\sim1.0~\mu$m which is best explained by a Lyman 1216-${\rm \AA}$ break at $z=7.23\pm0.05$ (Fig.~1).

\targname\ is distinct in the rest-frame UV when compared to any other object currently known at similar redshifts ($z>6$). 
Its luminosity falls between typical quasars and galaxies and it is quite red in colour (Extended Data Fig.~1), with a rest-frame 1450\,${\rm \AA}$ luminosity, $M_{1450}$, of $-23.2$ mag and  
continuum slope, 
$F_{\lambda}\propto \lambda^{\alpha_{\lambda}}$,
of \(\alpha_{\lambda}= {0.1\pm0.3}\) (see Methods).
This is the reddest continuum slope found among objects at similar redshifts, ($\alpha_{\lambda}\lesssim-1.5$,\cite{bouwens2014,selsing2016}) but is comparable to lower-redshift red quasars identified in the Sloan Digital Sky Survey (SDSS)\cite{sdss_dr12} (Fig.~1). 
\targname\ is also spatially unresolved in all {\it HST} bands (Extended Data Fig.~2) and is bright at rest-frame 3~\mum, detected with {\it Spitzer}/MIPS. 
These characteristics suggest that \targname\ is a distant, red quasar.

However, \targname\ is not detected in the extremely deep 2\,Ms X-ray map of the Chandra Deep Field North\cite{xue2016}. 
Even accounting for obscuration, 
we obtain an upper limit (99\% confidence level) on the X-ray luminosity of $L_X < 3.9\times10^{42}$ erg~s$^{-1}$ (see Methods).  This is several orders of magnitude lower than what would be predicted by assuming the correlation between $L_X$ and optical luminosity observed for other quasars (Extended Data Fig.~3).
\targname\ is therefore strikingly faint in 2--10\,keV X-rays, apparently 
in tension with its being a quasar.
The absence of strong, broad UV emission lines (Fig.~1) in addition to this unique X-ray faintness raises the possibility that \targname\ could instead be an extreme UV-compact star-forming object.

1-  and 3-millimeter (mm) observations were carried out with the NOrthern Extended Millimeter Array (NOEMA) between June 2020 and February 2021 (see Methods). 
The \cii\ {158~\mum} line was robustly detected at 17$\sigma$ peak intensity  at a redshift of $z=7.1899\pm0.0005$, consistent with the Lyman-break redshift. 
The underlying 1\,mm and 3\,mm continua are also detected at 16$\sigma$ and 3.9$\sigma$ respectively.
The sky positions of the 1 and 3 mm continua and emission line are consistent with the \emph{HST} source. The \cii\ line is spatially resolved with an effective radius of $r_{\rm\,e}=1.4$\,kpc. The 1\,mm continuum is unresolved with an upper limit of $r_{\rm\,e}\leq0.5\,$kpc, suggesting that the majority of the far-infrared (IR) emission is arising from a compact region of $\leq0.7\,$kpc$^{2}$. 
No close neighbors are detected in the mm line or continuum maps.  However, a second source (``ND1'') is seen in the 1\,mm continuum map that is undetected in the deep \textit{HST} images and can be best explained as a dusty companion $\sim$16~kpc from \targname\ (see Methods). 

The precise redshift determination and the rich multi-wavelength datasets in the GOODS-North field provide a unique opportunity to constrain the host galaxy properties separate from the UV-luminous core.  
Fits to the optical-to-mm spectral energy distribution (SED) 
yield an IR luminosity (rest-frame 8--1,000~$\mu$m) of $L_{\rm\,IR} = (1.2\pm0.6)\times10^{13}$ \lsun\ {(where \lsun is the luminosity of the Sun)} (Fig.~2). 
This corresponds to a star-formation rate (SFR) of $1,600\pm700\,M_{\odot}$\,yr$^{-1}$ after removing the potential contribution of the emission associated with the active galactic nucleus (AGN). 
Regardless of the interpretation, the host of \targname\ is the most vigorously star-forming galaxy at $z>7$ found to date\cite{wang2021}.  It has a SFR surface density of $\geq1,100\,M_{\odot}\,$yr$^{-1}\,$kpc$^{-2}$, which is at the Eddington limit for star-forming galaxies\cite{andrews2011}. 
Treating \targname\ as a UV-compact star-forming object instead of an AGN would increase the SFR estimate.
The dust in the host 
has a peak temperature of \td~$\sim80$\,K, warmer than typical high-redshift quasar hosts by a factor of $\sim1.5$.\cite{beelen2006}
The $L_{\rm\,[CII]}/L_{\rm\,IR}$ ratio also shows one of the lowest values so far seen. 
The high dust temperature and relatively faint \cii\ emission may be due to the maximal SFR surface density, as the strong radiation field produced by the intense starburst increases \td\ and decreases the abundance of singly-ionized carbon. 
Dust and gas masses are estimated at $M_{\rm\,dust}= 1.6 \times10^{8}$\,\msun\ and $M_{\rm\,gas}= 2.0\times10^{10}$\,\msun\, making \targname\ one of the most dust- and gas-rich systems known at $z>6$ (see Methods). 
The presence of the proximate ND1 galaxy is consistent with the high abundance of companion galaxies reported around luminous quasars at $z>6$.\cite{decarli2017}

The shape of the NIR--MIR SED of \targname---especially the excess emission at rest-frame 3~\mum---cannot easily be explained by emission from stars and ionized gas associated with star-formation activity alone (Fig.~2 and Extended Data Fig.~8).
Moreover, fitting a profile to the {\it HST} images provides a stringent upper limit for the effective radius of only 60\,pc for the UV emission (see Methods).  
If this compact emission were attributed to star-formation, the SFR surface density from the UV alone would reach $\gtrsim$5,000\,$M_{\odot}$\,yr$^{-1}$\,kpc$^{-2}$, two orders of magnitude higher than the UV-luminous compact galaxies reported at $z\sim2$--3 (Extended Data Fig.~7)\cite{barro2014}. 
Taken together, the properties of \targname\ favor the interpretation of the UV source as a red quasar. Its properties are
in excellent agreement with the transition phase of the evolutionary paradigm of super-massive black holes (SMBHs): a low-luminosity, dust-obscured quasar emerging in a vigorously star-bursting host. The detection of dust-obscured, super-Eddington accretion objects hosted by starburst galaxies has so far been reported up to $z=4.6$.\cite{tsai2018} \targname\ at $z=7.2$ is found at a cosmic time that is 500 million years earlier, close to the earliest SMBH known at $z=7.6$.\cite{wang2021} The AGN in \targname\ is two orders of magnitude fainter than its lower-$z$ analog at $z=4.6$, but with a host SFR $\sim$3 times higher, suggesting that \targname\ is experiencing an early stage of its transition phase.

\targname's extreme X-ray faintness is a strong indicator of the young age of the quasar.
Extrapolating the anti-correlation\cite{lusso2010} between X-ray luminosity and AGN Eddington ratio ($\lambda_{\rm\,Edd}$) 
to the X-ray upper limit of \targname, we obtain an Eddington ratio significantly ({5.5}$\sigma$) greater than unity and a black hole mass of only $M_{\rm\,BH}\sim10^{7}\,M_{\odot}$ (Fig.~3a, see also Methods). 
X-ray-weak quasars are found to be abundant among weak emission line quasars (rest-frame equivalent width of \civ\ EW(CIV) $< 16 {\rm\,\AA}$)\cite{luo2015,pu2020} with more powerful nuclear winds\cite{wu2011}. 
These trends can be explained by a scenario where the inner region of the accretion disk is strongly inflated to a substantial height due to the unusually high accretion, which blocks the nuclear ionizing continuum and the X-rays from reaching the broad line region and external observers\cite{luo2015}. 
\targname\ is indeed lacking the \civ\ line in our spectroscopy (EW(CIV) $<$  10${\rm\,\AA}$), consistent with this scenario. 

These observational results can be compared with cosmological semi-analytic models for progenitors of high-$z$ quasars \cite{valiante2016,ginolfi2019}. 
Among simulated merger histories, several progenitors with multi-wavelength properties similar to \targname\ indeed have \tcb{relatively low mass SMBHs ($M_{\rm\,BH}\sim10^{6.5-7.5}\, M_{\odot}$)}, but still reside in the most massive halos of $\sim10^{11.5-12.5}\, M_{\odot}$ at $z=7.2$ (Fig.~4).
These simulations show that all of these progenitors will evolve into optically-luminous blue quasars harboring a SMBH with $M_{\rm\,BH}>10^{8}\, M_{\odot}$ at $z=6.4$.
This indicates that \targname\ could be the direct progenitor of an optically-luminous quasar, 
although models do not rule out the possibility that \targname\ 
will fail to finalize its transition at later epochs because of possible mergers with other halos hosting more massive BHs. 
The simulations and recent observations also predict a tight correlation between $M_{\rm\,BH}$ and the X-ray luminosity normalized by infrared luminosity $L_{\rm\,IR,\,SF}$, confirming that the unique X-ray faintness of \targname\ corresponds to the regime of $M_{\rm\,BH} < 10^{8}\,M_{\odot}$ (Fig.~3b). 

Given the short-lived nature of a transitioning red quasar and the intrinsically low sky density of the quasar population\cite{glikman2012}, it is remarkable to find \targname\ within the 170\,arcmin$^2$ GOODS-North field. 
Assuming instead a total survey area of the entire \textit{HST} archive of nearly 3\,deg$^{2}$, the identification of \targname\ suggests a sky density of 0.33\,deg$^{-1}$ and a lower limit of 3.3$\times10^{-3}$ deg$^{-1}$ based on the Poisson uncertainty at the 99\% single-sided confidence level\cite{gehrels1986}. 
However, the quasar luminosity function (QLF) and the red quasar fraction at $z\sim6$ suggest a much lower predicted sky density of 6.8$\times10^{-4}$ deg$^{-1}$, even for less luminous red quasars similar to \targname\cite{matsuoka2018,kato2020}. 
A recent \textit{HST} study also reports a potentially higher density of less-luminous quasars at $z\sim8$ than the QLF at $z\sim6$.\cite{morishita2020} 
Together, these results may imply that the red and/or less-luminous quasar population is more common at $z>7$ than our understanding to date up to $z\sim6$.\cite{ni2020}  
We note in passing that classical colour selections for high-$z$ quasars in ground-based surveys would recover the identification of \targname\ (see Methods). 
This implies that these quasar populations could have been missed in previous surveys due to their faint nature in the mid-IR (MIR) and X-rays and in their rest-frame UV lines, 
that are here overcome by the uniquely deep and rich multi-wavelength datasets of the GOODS-North field. 
A systematic high-resolution, deep imaging survey in the optical--MIR bands may discover additional objects similar to \targname. Furthermore, follow-up spectroscopy of broad Balmer lines for $z>7$ objects will become possible with the launch of the {\it James Webb Space Telescope}. This will have the power to decisively determine whether the quasar classification is correct as well as determine how common such quasars truly are.
Even a non-detection of broad lines would imply intriguing conclusions, i.e.\
the existence of extraordinarily luminous and compact star-forming regions or stark differences between the first quasars and their descendants.

\input{_acknowledgement}
\input{_author_contribution}

\tcb{
\subsection{Competing interests}
The authors declare no competing interests. 
}

\tcb{
\subsection{Corresponding Author} 
Seiji Fujimoto (fujimoto@nbi.ku.dk)
}

\clearpage 
\newpage

\textbf{Figures}

\begin{figure*}[h]
\begin{center}
\includegraphics[angle=0,width=1.0\textwidth]{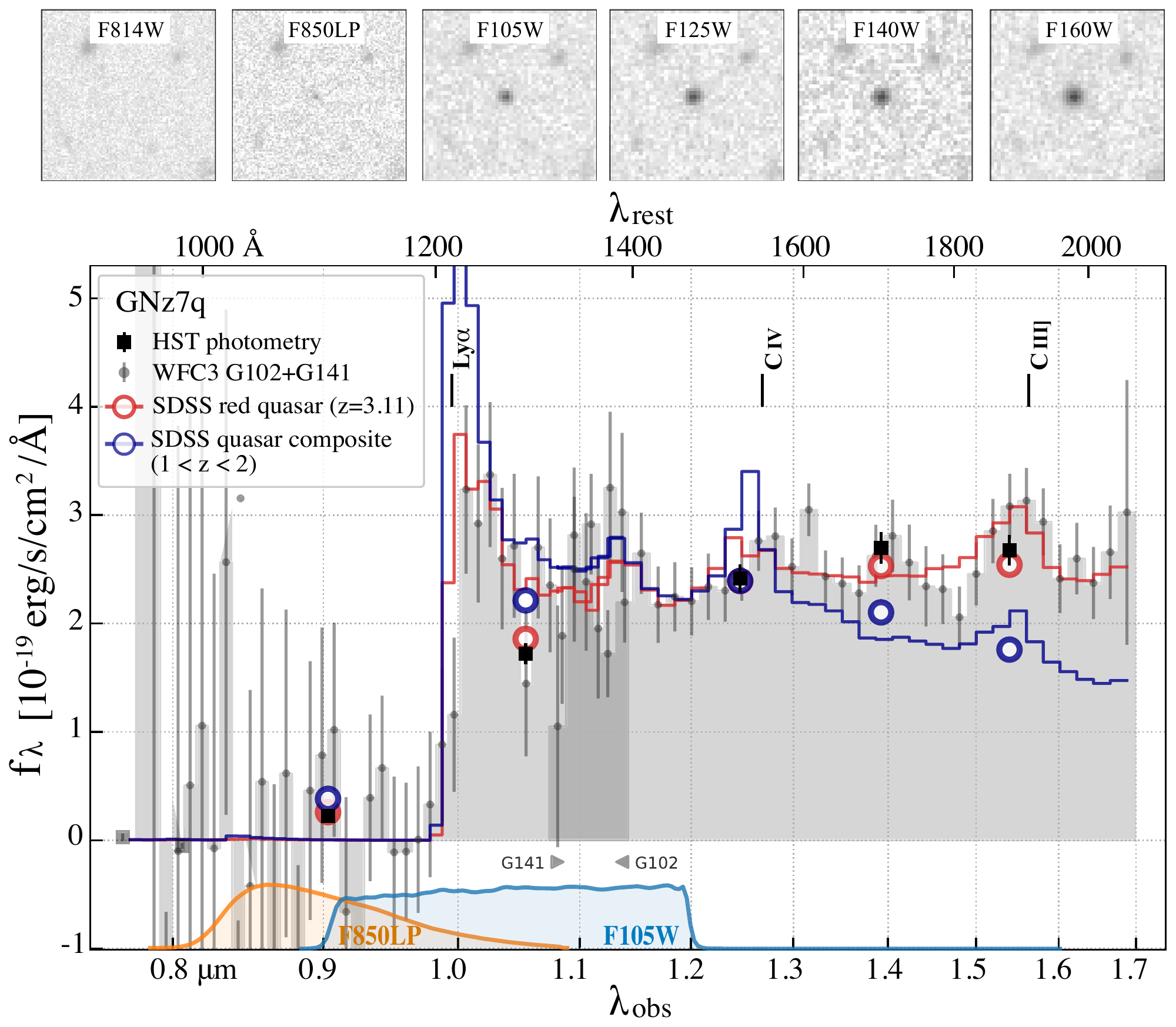}
\end{center}
\vspace{-0.6cm}
\caption{\small 
\tcb{
\textbf{$|$ \boldmath \emph{Hubble Space Telescope} near-infrared images and spectrum of \targname.} 
The spectrum and photometry show a strong Lyman break at $\lambda_\mathrm{obs}\sim1.0~\mu\mathrm{m}$.
The top panels show the \textit{HST} image cutouts ($5''\times5''$). 
The source is unresolved in all deep \textit{HST} images up to the reddest filter available at $1.6~\mu\mathrm{m}$ (WFC3/IR F160W).
In the bottom panel, the black squares and gray dots respectively show the broadband photometry and the slitless spectrum binned by a factor of 4 relative to the nominal pixel scale. 
The error bars denote 1$\sigma$ uncertainties. 
The labeled black bars indicate the expected wavelengths for the main emission lines based on the \cii\ 158-$\mu$m line redshift of \targname\ at $z=7.1899$.
The blue curve represents a composite spectrum of SDSS optically luminous blue quasars\cite{selsing2016} at $1<z<2$, while the red curve shows a red quasar at $z=3.11$ (\tcb{SDSS spec-6839-56425-146})\cite{sdss_dr12} whose FUV spectrum resembles that of \targname. 
Both of the lower-$z$ quasar spectra are shifted to $z=7.1899$, normalized at 1.2~$\mu$m, and binned to the same spectral resolution as the \targname\ spectrum. 
The large open circles show the quasar templates integrated through the \textit{HST} filter passbands.   
The bandpasses of the ACS/F850LP and WFC3/F105W filters shown at the bottom straddle the spectral break, explaining the faint detection in the former and the suppressed flux density relative to the continuum in the latter.}
\label{fig:grism}}
\end{figure*}
\newpage

\begin{figure*}[h]
\centering
\includegraphics[angle=0,width=1.0\textwidth]{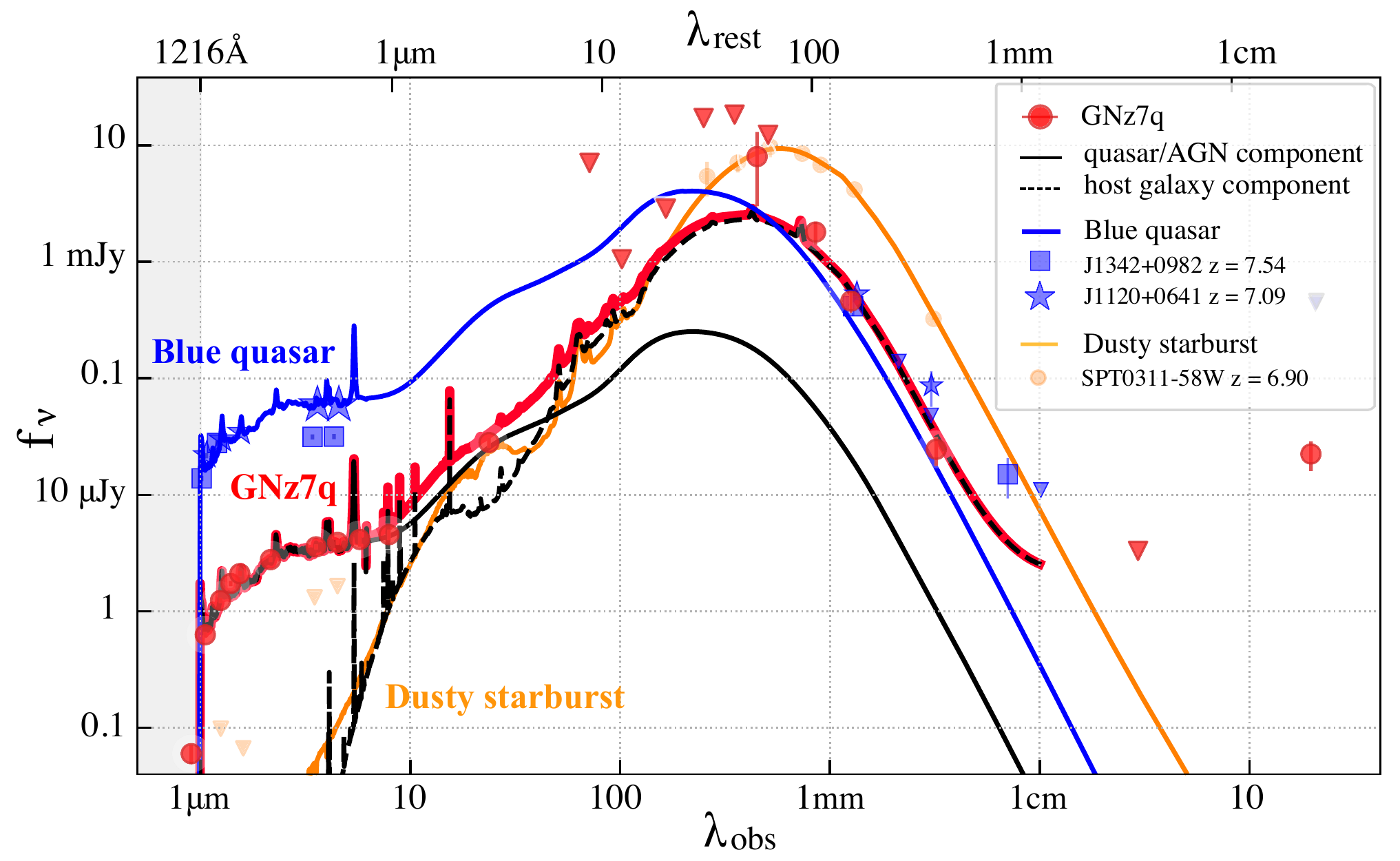}
\caption{\small 
\tcb{
\textbf{$|$ 
\boldmath
The spectral energy distribution of \targname\ from optical to radio wavelengths.}
Photometry is shown for data from \textit{HST} (0.8--1.6$\mu$m), \textit{Spitzer} (3.6--24$\mu$m), \textit{Herschel} (80--500$\mu$m), JCMT (450 and 850$\mu$m), NOEMA (1 and 3~mm) and VLA (3 and 20~cm) in the GOODS-North field (Extended Data Table~1).} 
Triangles indicate 3$\sigma$ upper limits. 
The sum of the best-fit quasar/AGN (black solid) and galaxy (black dashed) templates is shown as a red curve. 
The radio detection at 20~cm is consistent with the enormous implied SFR of the host galaxy (see Methods).
For comparison, we also show the SEDs of other source populations at similar redshifts: optically-luminous blue quasars at $z=7.54$ (J1342+0928\cite{banados2018}; blue squares) and $z=7.08$ (J1120+0641\cite{mortlock2011}; blue stars), and a dusty starburst at $z=6.90$ (SPT0311-58W\cite{marrone2018}; orange circles). 
The blue curve is drawn with the quasar/AGN template normalized to J1120+6410's rest-frame UV emission. 
The orange curve is the best-fit SED for SPT0311-58W, taken from the  literature\cite{marrone2018}. 
The SED of \targname\ falls between these two categories of the dusty starburst and the blue quasar, representing a transient phase between them.
\label{fig:full_sed}}
\end{figure*}

\newpage

\begin{figure*}[h]
\centering
\includegraphics[angle=0,width=1.0\textwidth]{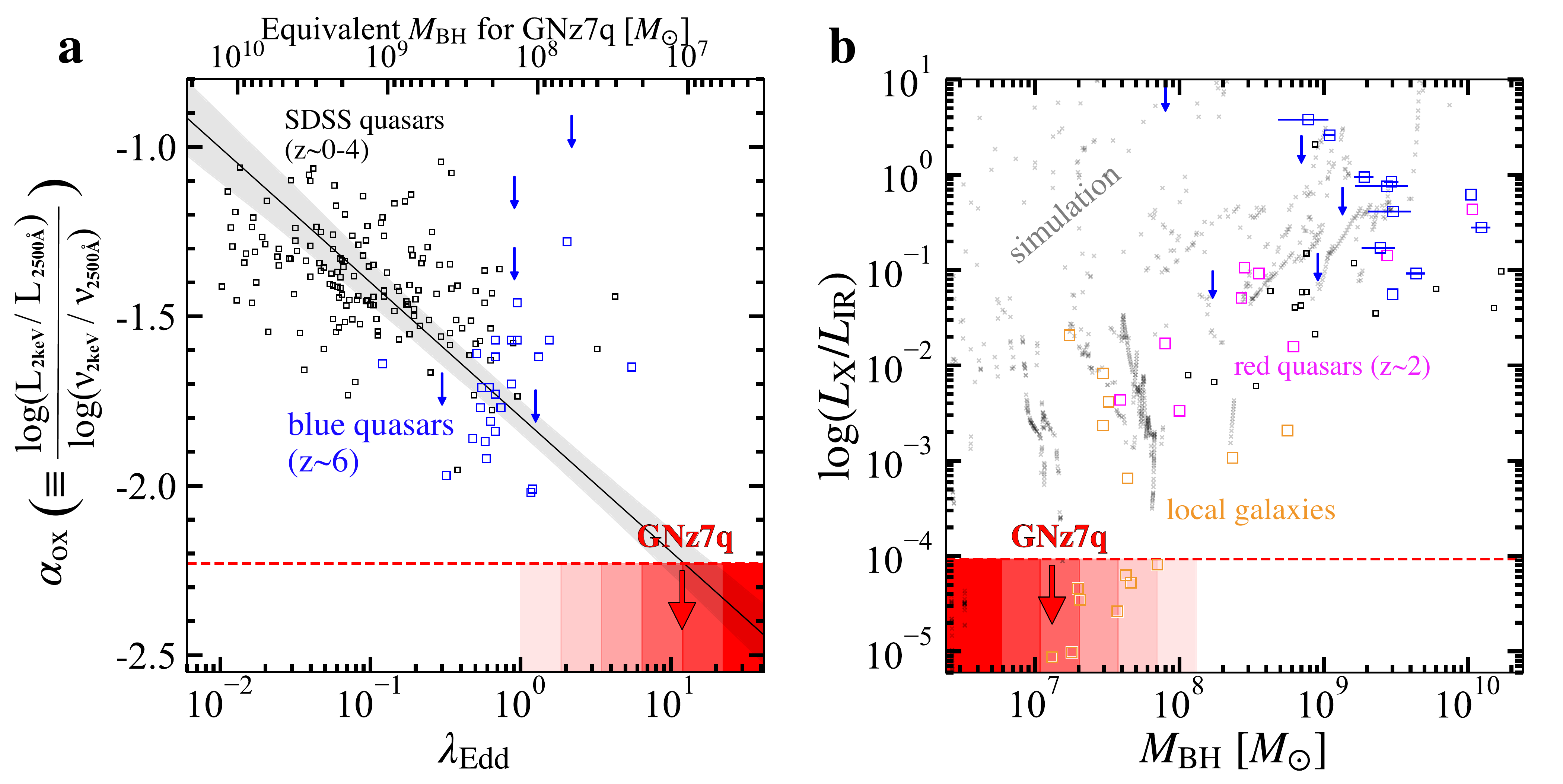}
\caption{\small 
\textbf{$|$  
The unique X-ray faintness of \targname.} 
The 2\,Ms deep \emph{Chandra} data\cite{xue2016} provide a stringent upper limit (red dashed line) for the X-ray luminosity, suggestive of a very high accretion rate ($\lambda_{\rm\,Edd}\gtrsim1$) onto a less massive black hole ($M_{\rm\,BH}\lesssim10^{8}\,M_{\odot}$). 
{\bf a.} Optical to X-ray spectral index $\alpha_{\rm\,ox}$ as a function of Eddington ratio ($\lambda_{\rm\,Edd}$). 
{The dust-corrected optical luminosity $L_{\rm 2,500}'$ is used in the upper limit estimate of $\alpha_{\rm\,ox}$ for \targname.} 
SDSS quasar measurements (black squares) and the best-fit relation (black line) with its $1\sigma$ confidence level (gray shaded region) are taken from the literature (see Methods).  
The upper horizontal axis shows the equivalent black hole mass for \targname\ as a function of $\lambda_{\rm\,Edd}$ based on its AGN bolometric luminosity from the UV to mm SED fitting (see Methods).
The red shaded region shows the $\lambda_{\rm\,Edd}$ regime of \targname\ extrapolated from the best-fit relation, where the shading becomes darker with increasing $\lambda_{\rm\,Edd}$. 
{\bf b.} X-ray luminosity ($L_{\rm\,X}$) normalized by $L_{\rm\,IR}$.  
We show other populations for comparison: {blue quasars at $z\sim6$ (blue squares), red quasars at $z\sim2$, and dusty starbursts at $z\sim0$ (orange squares)} taken from the literature (see Methods). 
Gray circles are plotted for simulated galaxies (see Methods) 
with AGN bolometric luminosity of $L_{\rm\,bol}>10^{42}$ erg~s$^{-1}$. 
The colour scale and the horizontal range of each red shaded region of $M_{\rm\,BH}$ corresponds to those of panel {\bf a}. 
\label{fig:xray}}
\end{figure*}
\newpage

\begin{figure*}[h]
\centering
\includegraphics[angle=0,width=1\textwidth]{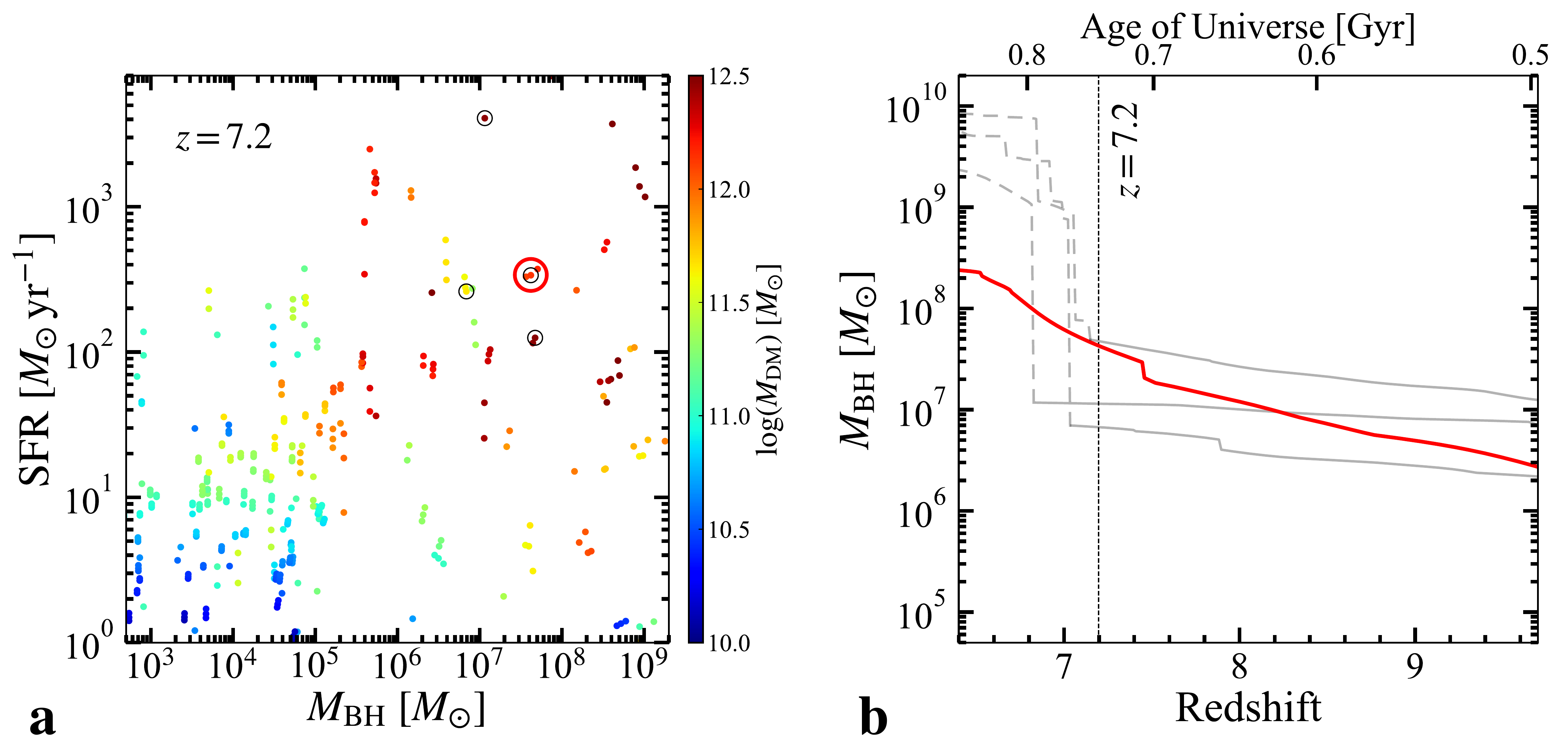}
\caption{\small \textbf{$|$  
SFR and $M_{\rm\,BH}$ relations for progenitors of luminous quasars in a cosmological simulation}. 
\textbf{a.} SFR and BH masses predicted by the semi-analytical model GQd\cite{valiante2016} for selected $z = 7.2$ progenitors of a luminous quasar at $z=6.4$. 
Each system is represented by a circle colour-coded by the dark matter halo mass ($M_{\rm DM}$).  
The four black circles mark systems that have X-ray, optical, and host galaxy properties similar to \targname\ (See Methods).
\textbf{b.}
BH mass assembly history for the systems marked with the black circles in panel {\bf a}.
At $z = 7.2$, one of them has already grown to $4.0\times10^{7}\,M_{\odot}$ (red circle in panel a) and thereafter it continues to grow by gas accretion and mergers with other BHs to form a SMBH of $M_{\rm\,BH}=2.5\times10^{8}\,M_{\odot}$ at $z=6.4$ (red line). 
The other systems also have relatively low-mass BHs, down to $\sim10^{6.5-7.5}\,M_{\odot}$ at $z=7.2$ (grey lines), but these systems undergo mergers with galaxies hosting more massive BHs.
As a result, these BHs are not the most massive progenitors of the final SMBH, which grows to $\sim10^{9.3-10}$ $M_{\odot}$ by $z=6.4$ through gas accretion and mergers with other BH progenitors (grey dashed line).
The black vertical line indicates the redshift of \targname. 
\label{fig:simulation}}
\end{figure*}
\newpage

\clearpage

\newpage 
\begin{methods}
In this paper, error values represent the $1\sigma$ uncertainty, where $\sigma$ denotes the root-mean-square or standard deviation;  upper limits are indicated at the 3$\sigma$ level; 
red symbols in figures denote \targname, unless otherwise specified.

\subsection{1. Cosmology} 

We adopt cosmological parameters measured by the
Planck mission\cite{planck2014}, i.e.\ a $\Lambda$ cold dark matter ($\Lambda$CDM) model with total matter, 
vacuum and baryonic densities in units of the critical density,
$\Omega_{\Lambda}=$ 0.76,  
$\Omega_{\rm m}$ = 0.24,  
$\Omega_{\rm b} =$ 0.04,  
and Hubble constant, $H_{0}=100$ $h$\,km\,$s^{-1}$\,Mpc$^{-1}$, 
with $h=$ 0.73 
Based on these parameters, we adopt the angular size distance of 5.32 kpc/arcsec at the source redshift of $z=7.2$ in this paper.

\subsection{2. Definition of quasar categories} 

In this paper, we make a distinction between a reddened Type 1 quasar and a Type 2 quasar, where we refer to the former as the red quasar.  
A Type 1 quasar is defined to have at least one broad emission line (FWHM $\gtrsim$ 1,000--2,000~km~s$^{-1}$) in the spectrum, while Type 2 quasars are defined by those that do not satisfy it, but has too bright narrow UV-optical or IR emission lines, X-ray, or radio continuum for a galaxy\cite{onoue2021}. In the classical AGN unification models, this difference is generally explained by different viewing angles toward the central accretion disk\cite{antonucci1993,urry1995}, where the observer’s line of sight in Type 2 quasars penetrates through optically-thick dusty material due to the nearly edge-on view of the dust torus that blocks the nuclear broad emission lines.  
In this context, red quasars are generally defined to have at least one broad emission line in the spectrum, distinguishing them from Type 2 quasars, but to be substantially reddened by dust. 
The extent of the dust reddening has been estimated a number of ways in the literature based on optical\cite{richards2003,glikman2012,kato2020}, optical-MIR\cite{ross2015, hamann2017}, NIR-radio\cite{glikman2004,urrutia2009}, and MIR\cite{lacy2007,glikman2013}. 
For example, the red quasar is characterized by the optical colour excess of $E(B-V)\gtrsim0.1$ in the optical colour selection\cite{richards2003,glikman2012,kato2020}. 
Following these definitions, 
\tcb{the SDSS quasar at $z=3.11$ (Fig.~1, SDSS spec-6839-56425-146) is classified as a red quasar with $E(B-V)$ = 0.13.\cite{krawczk2015}}
W2246-0526 is also classified as a red quasar that is a super-Eddington, extremely luminous, and dust-obscured quasar at $z=4.6$ with broad Mg{\sc ii} and C{\sc iv} emission lines and $E(B-V)$ = 5. \cite{tsai2018,diaz-santos2018,diaz-santos2021}
Because several sub-millimeter galaxies (SMGs; flux density at 850~$\mu\mathrm{m}$ $\gtrsim$ a few mJy) at $z\sim2$ are reported to have broad emission lines and $E(B-V)\sim0.4$, we regard these SMGs also as red quasars\cite{alexander2008} in this paper. 
Local (ultra) luminous infrared galaxies ((U)LIRGs) and high-redshift SMGs that are not confirmed to have broad emission lines are referred to as dusty starbursts, 
while we refer to the hot dust-obscured quasars confirmed with the broad emission lines\cite{finnerty2020} as red quasars in this paper.
We refer to Type 1 quasars from the literature that do not exceed
the red colour threshold of $E(B-V) > 0.1$ as blue quasars in this paper.

\subsection{3. Uniform processing of archival \textit{HST} and \textit{Spitzer} data} 

\targname\ was identified in a project nicknamed the Complete Hubble Archive for Galaxy Evolution (CHArGE). CHArGE aims to perform uniform processing and analysis of all archival {\it HST} and {\it Spitzer} data taken away from the Galactic midplane. 
\textit{HST} Advanced Camera for Surveys (ACS) optical and the Wide Field Camera 3 (WFC3) near-infrared and {\it Spitzer}/Infrared Array Camera (IRAC) observations covering \targname\ were carried out by a variety of large extra-galactic surveys and individual programs.  
\textit{HST} images were obtained in the F435W (6 exposures; 7.2~ks integration), F606W (23~exp.; 9.3~ks), F775W (35~exp.; 19~ks), F814W (42~exp.; 24~ks), F850LP (82~exp.; 37~ks), F105W (6~exp.; 3.3~ks), F125W (8~exp.; 4.4~ks), F140W (6~exp.; 1.2~ks), and F160W (8~exp.; 5.4~ks) filters. The IRAC channel 1 ($3.6~\mu\mathrm{m}$) and channel 2 ($4.5~\mu\mathrm{m}$) integrations are 345$\,$ks and 330$\,$ks, respectively.  We aligned all of the {\it HST} exposures to sources in the {\it GAIA} DR2 catalog\cite{gaia2018} and created final mosaics in a common pixel frame with 50\,mas and 100\,mas pixels for the ACS/WFC and WFC3/IR filters, respectively.  
We aligned the individual \textit{Spitzer} exposures to the same astrometric frame as the {\it HST} frame and generated final drizzled IRAC mosaics with $0.''5$ pixels.  Further details of the {\it HST} ({\it Spitzer}) image processing with the \texttt{grizli} (\texttt{golfir}) software will be presented in Kokorev (in prep.). 
In Fig.~1, we present the {\it HST} images of \targname.

Archival {\it HST} slitless spectroscopy of \targname\ is available with integration times 8.8 ks (12.7 ks) in the G102 (G141) grisms from HST General Observer (GO) program 13420 (11600).  Together the two grisms cover $0.8 < \lambda < 1.7~\mu$m without any gaps (Fig.~1).  The slitless spectra are reduced and extracted with the \texttt{grizli} software\cite{brammer2019}.

The {\it HST} broad-band images and slitless spectroscopy show an unambiguous continuum break at $\sim1.0$ $\mu$m. 
By using a SDSS red quasar template at $z=3.11$ (\tcb{SDSS spec-6839-56425-146})\cite{sdss_dr12,krawczk2015}, we obtain the rest-frame UV redshift at $z_{\rm UV}=7.23\pm0.05$. 
This is consistent with the \cii\ 158-$\mu$m line redshift of $z_{\rm [CII]}=7.1899\pm0.0005$ within the uncertainties (Section 4). 
We fit a power-law ($F_{\lambda} \propto \lambda^{\alpha{\lambda}}$) model to the rest-frame UV continuum of the G141 slitless spectrum at 1.10--1.66~\mum, excluding 1.50--1.60~\mum to avoid potential contribution from a \ciii$\lambda$1909 line and the noisy edge of the spectrum (Fig.~1),
and measure a best-fit UV continuum slope $\alpha_{\lambda}=0.1\pm0.3$. 
By subtracting the best-fit power-law continuum and optimizing the integration range of the spectrum, 
the broad Ly$\alpha$ and \ciii\ lines are tentatively detected at 3.7$\sigma$ and 3.4$\sigma$ levels, respectively.
Assuming a \civ\ line width of full-width-half-maximum (FWHM) $=$ 3,000~km~s$^{-1}$ typical among $z>6$ quasars from the literature\cite{shindler2020}, we derive a 3$\sigma$ upper limit of EW$_{\rm CIV} = 10\,{\rm \AA}$ from the grism spectrum.

In Extended Data Fig.~1, we compare the rest-frame UV properties of \targname\ with other populations at similar redshifts. 
The absolute UV luminosity at 1450 ${\rm \AA}$ ($M_{1450}$) is estimated from the best-fit power-law model. 
We find that the UV luminosity of \targname\ falls between typical quasars and galaxies in the literature\cite{shindler2020,inayoshi2020,yang2020,wang2021}, being $\sim$10 times fainter and brighter than typical quasars and galaxies, respectively, where several faint quasars and luminous galaxies have been reported\cite{bowler2017,schouws2021,matsuoka2016,matsuoka2017,matsuoka2018b,matsuoka2019,matsuoka2019b,onoue2019}. The UV continuum slope of \targname\ is redder than that of any other object in either comparison population. 

Extended Data Fig.~2 shows the rest-frame UV morphology in all {\it HST} WFC3/IR bands and the radial profile of \targname\ observed with {\it HST}/F125W. 
The 
instrumental point spread function (PSF) model can fully explain the rest-frame UV morphology / radial profile in all bands. The S\'ersic profile fitting with \texttt{galfit}\cite{peng2010} provides almost the same profile as the PSF model in all bands. 
\tcb{Based on the F125W band, we measure an effective radius of $r_{\rm e}=0.06\pm0.07$ pixel (pixel scale = $0.''06$) and adopt a 2$\sigma$ upper limit of $r_{\rm e}< 0.02$~pix $\simeq60$~pc.} 
We obtain similar results in the S\'ersic profile fitting to the other bands.

\subsection{3. NOEMA observations, data reduction, and measurements}

We observed \cii\ with band 3 (1~mm), and CO(7–6), CO(6-5), and \ci\ lines with band 1 (3~mm) of Institute Radio Astronomie Millim\'etrique (IRAM) NOrthern Extended Millimeter Array (NOEMA). The observations were carried out between 2020 June 17 and 2021 February 24 in various visits with the AC and D array configurations for the 1~mm and 3~mm observations, respectively, using 9--10 antennas. The data were processed in the standard manner with the pipeline using the latest version of the GILDAS software. We used CASA version 5.6 for the imaging\cite{mcmullin2007}. 

For the \cii\ observations, the upper side band (USB) of the 1~mm band receiver was tuned to 231.8 GHz in the first execution with the C configuration to cover the \cii\ line at the source redshift estimated from  the FUV Lyman continuum break (Fig.~1). After confirming the \cii\ line detection,  we tuned lower side band (LSB) with the A configuration to cover a wide frequency range for the continuum emission. 
For the 3~mm observations, the 3~mm band receiver was tuned to 97.7 GHz to cover CO(7-6) and \ci\ in the USB and CO(6-5) in the LSB. 
In both observations, 0851+202 served as band-pass phase calibrator. Additional targets 1300+580 and 1044+719 were used for the phase and amplitude calibrations. We calibrated the absolute flux scale against  
MWC349 whose flux is regularly monitored at NOEMA. 
We adopt conservative uncertainties on the absolute scale of 20\% and 10\% in the 1~mm and 3~mm observations, respectively.  
The total integration time on-source was 6.8 and 13.5 hours in the 1~mm and 3~mm bands, respectively. 

To maximize sensitivity, we used natural weighting for the imaging. 
The resulting 1~mm and 3~mm maps have synthesized beam FWHM of $0.''64\times0.''44$ and $4.''7\times4.''1$, with 1$\sigma$ sensitivities of 21 and 6.3~$\mu\mathrm{Jy}~\mathrm{beam}^{-1}$ for the continuum, and 0.28 and 0.17~$\mathrm{mJy}~\mathrm{beam}^{-1}$ for the line per 60~$\mathrm{km~s}^{-1}$ channel, respectively.
We produce the 1~mm and 3~mm continuum map from the all line-free channels, except for noisy channels around the central frequency channels of LSB and USB, and the line cubes with several velocity bins in the range of 40--60~$\mathrm{km~s}^{-1}$. 
The central wavelengths of the 1~mm and 3~mm continuum maps are 1.284~mm and 3.276~mm, respectively. 

The continuum is detected both from the 1~mm and 3~mm maps with the peak intensity at 16$\sigma$ and 3.9$\sigma$, respectively. 
The \cii\ line is robustly detected at $232.060\pm0.013$ GHz with  FWHM=$280\pm40$~km~s$^{-1}$ and 17$\sigma$ peak intensity in the velocity-integrated map. 
This provides a precise determination of the source redshift of  $z=7.1899\pm0.0005$. 
At this source redshift, the CO(7-6) line is also detected at 5.7$\sigma$ peak intensity in the velocity integrated map with FWHM of $770\pm230$ km s$^{-1}$. 
While the velocity-integrated CO(7-6) line shows a signal at a marginal significance level, we regard it as a tentative detection due to the difference in its line width from that of the \cii\ line beyond the 1$\sigma$ uncertainties. 
With optimised apertures, we estimate emission line luminosities of $L_{\rm [CII]}=$ $(1.1\pm0.3)\times10^{9}$ $L_{\odot}$ and $L_{\rm CO(7-6)}=$ $(1.3\pm0.7)\times10^{8}$ $L_{\odot}$. 
For area-integrated line intensities, we measure $\log L_{\rm [CII]}'$ $=(5.0\pm1.4)\times10^{9}$  K~km~s$^{-1}$~pc$^{2}$ and $\log L_{\rm CO}'$ $=(6.3\pm3.0)\times10^{9}$  K~km~s$^{-1}$~pc$^{2}$, respectively. 
CO(6-5) and \ci\ lines are not detected with 5$\sigma$ upper limits of $L_{\rm CO(6-5)}< 5.0 \times10^{7} L_{\odot}$  and $L_{\rm [CI]2-1}< 7.9 \times10^{7} L_{\odot}$. 
In Extended Data Fig.~\tcb{3 and 4}, 
we summarize the continuum, velocity-integrated maps, and the line spectra.
We do not identify a clear velocity gradient in the \cii\ line. 
We list the continuum flux densities and the line luminosities in Extended Data Table 1 and 2, which includes statistical errors based on the optimized apertures and the absolute flux uncertainty of 10--20\%, as discussed above. 

We measure the effective radius of the 1.3~mm continuum ($r_{\rm e, FIR}$) and the \cii\ line ($r_{\rm e, [CII]}$) emission from \targname\ (Extended Data Fig.~\tcb{5}) using the visibility-based fitting CASA task {\sc uvmodelfit}. 
We adopt the 2-dimensional elliptical Gaussian model and leave all parameters free in the fitting routine.
{We obtain a best-fit major axis radius of $r_{\rm [CII]}=0.''26\pm0.''04$ ($1.40\pm0.21$ kpc). 
On the other hand, the continuum fit did not converge and indicates that the continuum is not spatially resolved.
The image-based fitting CASA task {\sc imfit} provides consistent results for the spatially resolved \cii\ line and  unresolved continuum. 
We estimate an upper limit to the continuum size based on an approximate formula for the smallest resolvable scale ($\theta_{\rm min}$) sampled by the interferometric data
\cite{marti2012}:
\begin{eqnarray}
\theta_{\rm min} = \beta \left(\frac{\lambda_{\rm c}}{2 S/N^{2}}\right)^{1/4} \theta_{\rm beam}, 
\end{eqnarray}
where $\beta$ is a coefficient, taking a range of 0.5--1.0, $\lambda_{\rm c}$ is a probability cutoff, and $\theta_{\rm beam}$ is a synthesized beam FWHM. 
Based on the 2$\sigma$ cutoff in the same manner as previous studies\cite{fujimoto2017,franco2018,franco2020} and $\beta=1.0$, we obtain the upper limit in FWHM of $0.''18$ ($\sim0.''09$ in effective radius) for the 1.3~mm continuum, corresponding to  $r_{\rm FIR} \leq 0.48~\mathrm{kpc}$).
}
In Extended Data Fig.~\tcb{5}, we present the observed and residual maps and the best-fit $uv$ visibility plot for the continuum and \cii\ line emission.

A nearby object is detected (9.4$\sigma$) in the 1.3~mm continuum map $\sim3.''1$ northeast of \targname, corresponding to a projected separation of $\sim$16~kpc if the two sources are at the same redshift. 
We obtain the 1.3~mm band photometry of 247 $\pm$ 65 $\mu$Jy for this object within a $1.''0$ radius aperture. 
We do not identify any line features in either the 1~mm or 3~mm band spectra extracted at the position of the nearby object, nor do we identify any counterpart in the deep multi-wavelength imaging data from \textit{HST} and \textit{Spitzer}. We refer to this NIR-dark nearby object as ``ND1'' and consider two interpretations for it being either a chance projection of a foreground galaxy or a companion object associated with \targname. 
Such a population of NIR-dark galaxies has been recently reported, perhaps representing the most massive galaxies at $z\sim$3--5.\cite{wang2019} 
However, given the low surface density of such galaxies, 
there is a very low probability to identify a ND1-like galaxy by chance within a small radius ($\sim3.''1$) from another particularly unique object like \targname. 
In fact, similarly NIR-dark galaxies have been recently identified at the same redshift as nearby massive galaxies at $z\sim7$.\cite{fudamoto2021} 
Therefore, we speculate that ND1 is a companion system associated with \targname\ which could eventually merge into a single system. 
Note that if we assume a typical modified blackbody (MBB) with $T_{\rm d}=$ 35 K and $\beta_{\rm d}=1.8$ for the FIR SED of ND1 and a typical $L_{\rm [CII]}/L_{\rm IR}$ ratio of 10$^{-3}$, 
the expected \cii\ line luminosity agrees with the non-detection of the \cii\ line from ND1 based on the current 1~mm data. This FIR SED for ND1 is also consistent with the non-detection in the 3~mm continuum due to its faintness and the cosmic microwave background (CMB) effect\cite{dacunha2015,jin2019,cortzen2020}. 
ND1 was subtracted before the {\sc uvmodelfit} analysis, and thus the presence of ND1 does not affect our size measurements of the primary target. 

\subsection{4. Multi-band photometry}  

The optical and NIR photometry are mainly obtained from the CHArGE data sets. 
We further obtain multi-band photometry of \targname\ from X-ray to radio from the literature. Given the spatial resolution of each instrument, we adopt the cross-match radii in the range from $1.''0$ to $4.''0$ with the public catalogs in the literature. 
We identify that \targname\ is robustly detected also at 2.2 $\mu$m (Subaru/Multi-Object
InfraRed Camera and Spectrograph (MOIRCS)), 24 $\mu$m ({\it Spitzer}/MIPS), 850 $\mu$m ((Submillimetre Common-User Bolometer
Array 2 (SCUBA2)), and 20 cm (JVLA) \cite{kajisawa2011,magnelli2011,cowie2017, owen2018}. 
For other data at 100--500 $\mu$m ({\it Herschel}/Photodetector Array Camera
and Spectrometer (PACS) and the Spectral and Photometric Imaging Receiver(SPIRE)) and 3 cm (JVLA)\cite{liu2018,oliver2012,murphy2017}, we do not identify any counterparts. 
{We measure the actual value at the source position in the JVLA 3~cm map, while we conservatively adopt $3\sigma$ upper limits for the {\it Herchel} maps from the instrumental and confusion sensitivities. 
}
We summarize the multi-band photometry and the literature in Extended Data Table~1. 
Note that we identify three nearby objects in multiple {\it HST} maps within a radius of $\sim2''$ (see top panels of Fig.~1). They are all detected in the F814W band and thus regarded as lower-$z$ objects, instead of companion galaxies at $z\sim7.2$. 
Their contribution to the {\it Spitzer} photometry of \targname\ is subtracted or negligible in our analysis. 
They are resolved in the IRAC images with de-blended 4.5$\mu$m flux densities of $0.08\pm0.03$, $0.08\pm0.02$ and $0.19\pm0.02$ $\mu$Jy, clockwise from the lower left.
Besides, they are bluer than \targname, where galaxy templates that include re-emitted dust emission in the MIR have 24$\mu$m / 4.5 $\mu$m ratios between 0.5 and 10. Thus, the faint neighboring sources together could contribute a maximum of 10\%, likely much less, of the unresolved 24-$\mu$m source at the position of \targname. 

For the SCUBA2 450 $\mu$m photometry, we produced the continuum map by utilizing the all archive data existing in this field. 
We reduced the data and performed the imaging in the standard manner with the pipeline based on the \texttt {starlink} software. 
We identify a tentative (2.6$\sigma$) detection in the 450 $\mu$m map including the flux-boosting correction based on the previous studies\cite{geach2017}. 
Because we identify the nearby object of ND1 in the NOEMA 1~mm map whose offset is smaller than the SCUBA2 beam, we subtract the potential contribution from ND1 in the 450~$\mu$m photometry. We estimate the expected 450-$\mu$m flux density of ND1 by assuming the typical MBB ($T_{\rm d}=$ 35 K and $\beta_{\rm d}=1.8$) at $z=7.2$. In the same manner, we also subtract the contribution from ND1 in the 850~$\mu$m photometry.  
We then obtain the 450 and 850~$\mu$m photometry of $8.0\pm5.0$ and $1.80\pm0.39$ mJy, respectively, that are also listed in Extended Data Table~1. Note that we confirm the consistency between the results with and without the tentative detection at the 450 $\mu$m band in the following analysis. 
Note that the contribution of the nearby faint {\it HST} objects to these FIR band photometry should be negligible due to the absence of the detection in the deep NOEMA 1~mm map. 
Although ND1 is detected in the deep NOEMA 1~mm map, a higher resolution 3~mm map with the briggs weighting (robust $=0.0$) whose beam size is smaller than the offset of ND1 from \targname\ shows $-4\pm9\mu$Jy at the position of ND1. Thus, the 3~mm photometry of \targname\ is not affected by ND1. 

In X-rays, there are zero events in the relevant pixel of \targname, even with the \textit{Chandra} 2~Ms integration, some of the deepest X-ray data ever taken\cite{xue2016}. 
The \textit{Chandra} 2~Ms data have 0.171 mean background counts in the full band. 
Based on the continuous Poisson distribution with the mean of 0.171, we compute an upper limit on the net counts of $<$ 1.1 at the 99\% confidence level. 
Assuming an average photon index $\Gamma=2.0$ obtained among high-redshift quasars up to $z=7.5$,\cite{nanni2017,banados2018,vito2018} 
we use the online Portable Interactive Multi-Mission Simulator (see code availability) and 
estimate upper limits for the X-ray luminosity to be $L_{\rm 2keV} < 5.1\times10^{24}$ erg~s$^{-1}$~Hz$^{-1}$ at 2 keV and $L_{\rm X} < 3.9\times10^{42}$ erg~s$^{-1}$ at 2--10 keV, including a correction for the Galactic absorption in this direction with the hydrogen column density of $N_{\rm H}=9.64\times10^{19}$~cm$^{-2}$.\cite{hi4pi2016} 
Note that the upper limit estimate depends on the choice of $\Gamma$. The typical range of spectral indices for X-ray AGN at this redshift is $\Gamma=$ 1.7--2.3.\cite{wang2021b} Assuming even an extremely soft spectrum of $\Gamma=2.3$ increases the luminosity upper limit by $\sim30\%$, still well below the expected value.

Following the definition of the optical to X-ray spectral index 
\begin{eqnarray}
\alpha_{\rm ox}\equiv\frac{\log(L_{\rm 2keV}/L_{\rm 2,500})}{\log(\nu_{2keV}/\nu_{\rm 2,500})}, 
\end{eqnarray}
we then obtain the upper limit of $\alpha_{\rm ox} < -2.23$. 
Note that here we use the dust corrected $L'_{\rm 2,500}$ value for $L_{\rm 2,500}$. 
In Extended Data Fig.~\tcb{6}, we show the tight correlation between $\alpha_{\rm ox}$ and $L_{\rm 2,500}$ previously observed for local and high-$z$ quasars\cite{shemmer2006, lusso2010, lusso2016, nanni2017, vito2019}. The stringent upper limit of the X-ray luminosity makes \targname\ deviate from this correlation by more than 5$\sigma$. 
In Fig.~3a, we show another tight correlation between $\alpha_{\rm ox}$ and $\lambda_{\rm Edd}$ known to exist among local and high-$z$ quasars\cite{lusso2010, lusso2016, nanni2017, vito2019,chiaraluce2018, zou2020}. 
The black line shows the best-fit relation estimated in Lusso et al. (2010), and the gray shaded region represents the 68th percentile of the relation evaluated by propagation from the 1$\sigma$ uncertainties of the parameters that define the best-fit relation. 
This best-fit relation yields $\log(\lambda_{\rm Edd}) = 1.1 \pm 0.2$ at the upper limit of $\alpha_{\rm ox}=-2.23$. This indicates that \targname\ has $\log(\lambda_{\rm Edd}) > 0$ at the $5.5\sigma$ level, and if $\alpha_{\rm ox}$ is much smaller than $-2.23$, the significance level could be much increased.  
In Fig.~3b, the values of $L_{\rm X}/L_{\rm IR,\,SF}$ and $M_{\rm BH}$ for other populations are taken from the literature\cite{nanni2017,vito2019,kim2019,iwasawa2011,veilleux2009,alexander2008}. 
We list the upper limits of the X-ray luminosity and $\alpha_{\rm OX}$ in the Extended Data Table~2. 

Note that it is difficult to explain the weak X-ray flux with selective absorption. This is because the redshift being so great means that we are observing very hard X-rays in the rest-frame. To extinguish the X-rays by a factor of 10 in the 2--10\,keV observed frame requires a column density that is Compton-thick. This corresponds to a typical dust column of $A_{\rm V}\gtrsim1,000$, which is orders of magnitude larger than our estimate of $A_{\rm V,\,qso}=0.3\pm0.1$ based on the rest-frame UV, optical, and NIR emission (Section~6). Furthermore, even if we were to assume that all of the gas was highly-ionised and the dust sublimated, so as to have no effect at optical wavelengths, the electron scattering depth would still be substantial, extinguishing the UV light by the same factor.
Therefore, the uniquely faint X-ray limits of \targname\ suggests either intrinsically weak X-ray emission, or Compton-thick material that covers only the inner part of the accretion disk and causes different extinction between the X-ray and the UV--NIR emission. The latter is aligned with the scenario that the inner region of the accretion disk is strongly inflated to a substantial height due to the high accretion, a so-called slim disk scenario\cite{luo2015,ni2018}.

\subsection{5. Interpretations for \targname} 

Based on the multi-wavelength observation results, there are two possible interpretations for \targname:  a Type 1 red quasar or a very compact UV-luminous starburst region.
The quasar interpretation is supported by the compact and luminous UV emission and the bright detection at rest-frame 3~\mum\, (MIPS 24~\mum). The starburst interpretation is motivated by the stringent upper limit on the X-ray luminosity and the absence of the clear detection of broad FUV emission lines between Ly$\alpha$ and \ciii$\lambda$1909.

In Extended Data Fig.~\tcb{7}, we compare the rest-frame UV size and luminosity of \targname\ with those of galaxies at $z>5.5$ and UV-compact galaxies at $2 \lesssim z \lesssim 3$\cite{shibuya2015,bowler2017,barro2014,marques2020}. 
For a more direct comparison, we show the observed UV luminosity without any dust correction. 
If the UV emission of \targname\ is interpreted as arising from hot, recently-formed stars, we find that the implied SFR surface density reaches $\gtrsim$ 5,000 $M_{\odot}$~yr$^{-1}$~kpc$^{-2}$, exceeding that of even the UV-compact galaxies by two orders of magnitude; any dust correction will make this more extreme still.
Such a luminous and compact object can more reasonably be explained by emission from a hot accretion disk in an active nucleus. 
In this case, 
the prominent FUV emission and its red continuum slope suggest that \targname\ is a Type 1 red quasar. 
The full SED analysis (Section~6) indicates an optical colour excess $E(B-V)=0.11\pm0.03$, which would satisfy the colour threshold for red quasars used in the literature\cite{richards2003,glikman2012,kato2020}.

Extended Data Fig.~\tcb{8} shows NIR--MIR SED and colour properties of \targname, galaxies, and both local and high-redshift quasars. The most striking feature of the \targname\ SED is the red color between the observed \textit{Spitzer} IRAC 8~\mum\ and MIPS 24~\mum\ band-passes, which probe rest-frame $\sim$1 and 3~\mum, respectively, at $z=7.2$.
A galaxy stellar population\cite{conroy2009, conroy2010} fit to the \targname\ photometry at $\lambda_\mathrm{obs} < 10~\mu\mathrm{m}$ is shown in the dark blue curve in Extended Data Fig.~\tcb{8}, along with additional stellar population templates that generally span the galaxy color space\cite{brammer2008} and that include highly obscured starbursts\cite{polletta2007}.  Also shown is the average MIR spectrum of local quasars\cite{glikman2006} and a collection of broad-band SEDs of quasars at $z\sim6$\cite{leipski2014}.  
The flux enhancement in the the MIPS 24~\mum\ band cannot  be explained by reasonable galaxy stellar population models, which are relatively blue between rest-frame wavelengths 1--3~\mum\ largely independent of star formation history and dust attenuation (i.e., the ``1.6~\mum'' bump feature).  However, the MIR colors of \targname\ are exactly consistent with the typical SEDs of quasars and active nuclei, where the rest-frame 3~\mum\ emission is thought to arise from hot dust associated with the nucleus\cite{nenkova2008, leja2018}.

Although the broad UV emission lines are not clearly detected (Fig.~1), 
a notable group of quasars with exceptionally weak UV broad emission lines have been identified at lower redshifts\cite{diamond-stanic2009,andika2020}. 
Moreover, the X-ray observations show that these weak emission line quasars also have weak X-ray emission with harder X-ray spectra and high-ionization lines that are more blue-shifted than those of normal quasars\cite{wu2011,wu2012,vito2021}. 
This suggests that the X-ray absorption is due to Compton-thick material covering the inner part of the accretion disk, consistent with the slim disk scenario (Section~4). 
A similar implication of the small, Compton-thick absorber has also been obtained from the weak X-ray emission in one of the nearest quasars embedded in the ULIRG-class dusty starburst Mrk~231,\cite{gallagher2002,braito2004} which also shows weak high-ionization lines and a very high nuclear outflow velocity of $\sim$8,000 km~s$^{-1}$.\cite{lipari1994,veilleux2016}  
Therefore, the extremely faint X-ray luminosity and the lack of broad UV emission lines do not necessarily contradict the quasar interpretation. 
In summary, the combined multi-wavelength properties favor the quasar interpretation of \targname, 
although we cannot rule out the possibility that we are witnessing an emergence of the extremely compact and luminous star-forming region that is without precedent in either the local or distant universe.

\subsection{6. UV--millimeter SED fitting}

We fit the available photometry from UV to mm wavelengths with the SED-fitting {code \sc{Stardust}}\cite{kokorev2021}, which models the emission from stars, active galactic nucleus (AGN),  and infrared emission arising from dust heated by star formation. 
The method does not rely on energy balance assumptions, and rather fits an independent linear combination of templates. 
To produce a fit we utilise the Draine \& Li (2007)\cite{draine2007} (hereafter DL07) models to fit the dust reprocessed stellar light, the AGN templates of Mullaney et al (2012)\cite{mullaney2012} as well as stellar and a quasar UV to NIR templates adopted from EAZY and Shen et al. (2016)\cite{shen2016}, respectively. For the quasar component, we include the SMC dust attenuation law to reproduce the red colours of \targname\ in the UV and NIR.

The SED-fitting yields the basic global UV--IR properties of the source: the observed ($L_{\rm 2,500}$) and dust-corrected ($L'_{\rm 2,500}$) monochromatic rest-frame optical luminosity, the dust attenuation for the quasar ($A_{\rm V,qso}$) and the host galaxy ($A_{\rm V,host}$), the total infrared luminosity (\lir) along with the relative contribution of the AGN (\lira) and of the star formation (\lirs) in the IR output of our target, the dust mass (\md), and the intensity of the mean radiation field ($\langle U \rangle \propto$ \lir/\md) of the source. The \lir, \lirs\ and \lira\ are estimated by integrating the corresponding best-fit components over  8--1,000~$\mu\mathrm{m}$. 
The bolometric luminosity of the AGN components that dominate the rest-frame FUV--NIR (observed 1--24\,$\mu\mathrm{m}$) is $L_\mathrm{bol} = 1.7 \pm 0.1 \times 10^{46}~\mathrm{erg~s}^{-1}$, where the small uncertainty is the random normalization uncertainty of the fixed templates constrained by the high S/N \textit{HST} and 24~$\mu\mathrm{m}$ detections.  The systematic uncertainty due to the relatively unconstrained MIR SED shape is somewhat larger.  For example, if we rather adopt the ``high luminosity'' MIR AGN template from Mullaney et al (2012)\cite{mullaney2012} as opposed to the ``low luminosity'' template favored by the full fit, we obtain $L_\mathrm{bol} = 1.3 \pm 0.1 \times 10^{46}~\mathrm{erg~s}^{-1}$.

For completeness and to facilitate comparison with the literature, we also fit the $\lambda_{\rm rest} >$ 50 $\mu$m data points with a MBB fit with a fixed $\beta_{\rm d}$=1.8 in order to infer a luminosity-weighted dust temperature (\td) and $L_{\rm IR,\,SF}$. 
We obtain the consistent $L_{\rm IR,\,SF}$ estimates from both fitting approaches. 
We use the average value of the $L_{\rm IR,\,SF}$ estimates and evaluate its uncertainty from the difference of the both estimates from the average. 
During the fitting process of both the multi-component analysis and of the MBB modeling, we have taken into account CMB effects that could be non-negligible at this redshift, especially in the mm regime\cite{dacunha2015,jin2019,cortzen2020}. 
The AGN/SF decomposition of the IR emission allows for an independent estimate of the obscured star-formation rate of the source. 
Using the conversion of Murphy et al. (2011)\cite{murphy2011} {based on the Kroupa initial mass function (IMF)\cite{kroupa2001},} the inferred \lirs\ corresponds to an SFR of of 1,600 $\pm$ 700 $M_{\odot}$~yr$^{-1}$, where the uncertainty of $L_{\rm IR, SF}$ is propagated to the uncertainty. 
Note that if we assume that \targname\ is rather a compact star-forming region, the obscured and un-obscured SFRs can be derived from $L_{\rm IR}$ and the observed UV luminosity, respectively. Using the same conversion, we obtain an even higher estimate of 1,900 $\pm$ 900 $M_{\odot}$~yr$^{-1}$, but we adopt the former estimate throughout this paper based on the discussion in Section 5. 
With the upper limit of the rest-frame FIR emitting region from the NOEMA 1~mm data, we also obtain a lower limit of the SFR density at $\geq$ {1,100} $M_{\odot}$~yr$^{-1}$~kpc$^{-2}$, which is comparable to its maximal value predicted from the balance between the radiation pressure from the star-formation and the self gravitation\cite{andrews2011,simpson2015}.   
Note that DL07 and MBB models assume the optically thin dust emission, while the compactness of the FIR-emitting region in \targname\ could in fact require an optically-thick treatment. With the optically-thick assumption, $L_{\rm IR,\,SF}$ and subsequently the SFR estimate will be increased by $\sim10$\% \cite{cortzen2020}.

Note that the uniquely deep X-ray data and general $L_{\rm X}$ and SFR relations\cite{lehmer2016} expect the X-ray detection 
from such an enormous SFR of the host galaxy, while the non-detection of X-ray could be explained by the metallicity dependence of the $L_{\rm X}$ and SFR relation: $L_{\rm X}$ at a given SFR decreases in high-metallicity systems\cite{fornasini2019}. Based on the recent calibration of Fornasini et al. (2020)\cite{fornasini2020}, we find that at least the solar value is required for the gas-phase metallicity of the host galaxy to meet the X-ray upper limit (Section~4). 
Since the SFR estimate for the host galaxy of \targname\ is not much changed regardless of its interpretation, the non-detection of X-ray thus suggests that the host galaxy has experienced a rapid metal enrichment and reached the solar metallicity even at $z=7.19$.
This is consistent with recent ALMA results for a similarly distant quasar at $z=7.54$ that the ISM gas-phase metallicity of the host galaxy is comparable to the solar value via FIR line diagnostics\cite{novak2019}.

\subsection{7. M$_{\rm gas}$ and M$_{\rm dyn}$ estimates}

We obtain molecular gas mass $M_{\rm gas}$ estimates from five empirical calibrations; the metallicity-dependent $\delta_{\rm GDR}-$method\cite{magdis2012} ($M_{\rm gas (dust)}$), the monochromatic Rayleigh-Jeans (RJ) dust continuum approach\cite{scoville2016} ($M_{\rm gas (RJ)}$), the CO line luminosity, the \cii\ line luminosity\cite{zanella2018} ($M_{\rm gas ([CII])}$), and the \ci\ line luminosity\cite{crocker2019} ($M_{\rm gas ([CI])}$). 
First, following the method in the previous studies \cite{magdis2012}, we convert the inferred \md\ to \mgas\ adopting a typical gas-to-dust ratio at solar metallicity, $\delta_{\rm GDR}$=92. This yields $M_{\rm gas (dust)}/$\msun = (1.4 $\pm$ 1.0) $\times10^{10}$  
which agrees well with the estimate inferred from the RJ approach, $M_{\rm gas (RJ)}/$\msun = (2.0 $\pm$ 1.2) $\times10^{10}$. For the latter we adopted a RJ luminosity-to-mass ratio $\alpha_{\nu}\equiv L_{\nu 850\mu {\rm m}}/M_{\rm gas}
= 6.7 \times 10^{19} \,{\rm erg~s^{-1}} Hz^{-1} M_{\odot}^{-1}$ calibrated with star-forming galaxies including local spirals and $z\sim2$ SMGs following Scoville et al. (2016)\cite{scoville2016}. 
Moving to the line tracers, we first estimate the area-integrated CO(1-0) intensity of $L_{\rm CO(1-0)}'=$ $(1.3\pm0.6)\times10^{10}$ Jy~km~s$^{-1}$~pc$^{2}$ from the CO(7-6) line detection, assuming $L'_{\rm CO(1-0)'}=$1.5$\times L'_{\rm CO(7-6)}$ estimated in dusty starburst galaxies at $z>6$ in the literature\cite{riechers2013,strandet2017}. 
We then estimated $(M_{\rm gas (CO)}/$\msun$)$= (5.0 $\pm$ 3.0) $\times10^{10}$ assuming $\alpha_{\rm CO}\equiv M_{\rm gas}/L_{\rm CO}'=4.6 $ $M_{\odot}$ 
(K~km~s$^{-1}$~pc$^{2}$)$^{-1}$ \cite{solomon1987}.   
To convert the \cii\ line to \mgas, we use the  $\alpha_{\rm [CII]}\equiv M_{\rm gas}/L_{\rm [CII]} = 22\, M_{\odot}/L_{\odot}$ conversion factor,  as calibrated on starburst galaxies at $z\sim2-6$ by \cite{zanella2018} and, adopting a 0.2~dex uncertainty, we estimate $M_{\rm gas (CII)} = 3.2\pm1.8\times\,10^{10}~M_\odot$. Finally, we consider the measured 3$\sigma$ upper limit of the \ci\ line that for   $\alpha_{\rm [CI](2-1)}\equiv M_{\rm gas}/L_{\rm [CI](2-1)} = 34$ $M_{\odot}$ (K~km~s$^{-1}$~pc$^{2}$)$^{-1}$ and a 0.32-dex uncertainty\cite{crocker2019, valentino2018}, we obtain an upper limit of $M_{\rm gas (CI)}/$\msun $<1.6\times\,10^{11}$. 
These independent \mgas\ estimates are in excellent agreement with each other within the uncertainties. For the purposes of this work, we determine $M_{\rm gas}=$ (2.0 $\pm$ 1.2) $\times10^{10}\,M_{\odot}$ by adopting the median value among $M_{\rm gas(dust)}$, $M_{\rm gas(CO)}$ and $M_{\rm gas([CII])}$. 
Note that the $M_{\rm gas(CO)}$ and $M_{\rm gas(dust)}$ estimates could be decreased by factors of several based on another assumptions of $\delta_{\rm GDR}=30$ and $\alpha_{\rm CO}=0.8$ in the super-solar metallicity case.

We estimate the dynamical mass of the system $M_{\rm dyn}$ from the \cii\ line results. 
Given the absence of the clear velocity gradient in \targname, 
we interpret \targname's host as a dispersion-dominated system and assume a virialized body with a radius $R$ kpc and one-dimensional velocity dispersion $\sigma$ km~s$^{-1}$, which yields\cite{bothwell2013} 
\begin{equation}
M_{\rm dyn} \,\, [M_{\odot}] = 1.56\times 10^{6} \sigma^{2} R .  
\end{equation}
For consistency with previous studies\cite{wang2013,decarli2018,izumi2018}, 
we use $r_{\rm e, [CII]}$ for $R$ after applying a correction factor of 1.5 to recover the contribution of the diffuse emission. 
We then obtain $M_{\rm dyn}=$ (4.5 $\pm$ 0.9) $\times10^{10}\,M_{\odot}$ which satisfies the requirement that it should be larger than the $M_{\rm gas}$ estimate. 
Because of the negligible dark matter contribution to $M_{\rm dyn}$ within a compact scale of 1.5$\times r_{\rm e, [CII]}$, we subtract $M_{\rm gas}$ from $M_{\rm dyn}$ and derive $M_{\rm star}$ of (2.5 $\pm$ 1.4) $\times10^{10}$ \msun.  
If we apply a typical dust-to-stellar mass ratio of 0.01 to $M_{\rm dust}$, we obtain another estimate of $M_{\rm star}$ of (1.6 $\pm$ 1.1) $\times10^{10}$ $M_{\odot}$ which is consistent with the above $M_{\rm star}$ estimate within the uncertainties.

\subsection{8. Comparison with other populations }

Compared to Type 2 quasars that are almost completely obscured in the UV/optical due to the nearly edge-on view of the dust torus, the reddening of red quasars is more moderate than that of Type 2 quasars, where the sight lines to red quasars may graze the dusty material surrounding the accretion disk. 
In this context, red quasars are thought to represent an early phase of the quasar life cycle: an obscured phase before the energy output from radiation, winds, and/or jets from the AGN and central star formation expels the obscuring material and transitions to an unobscured blue quasar. This is consistent with hydrodynamical simulations of galaxy mergers and quasar feeding\cite{hopkins2008}, and supportive observational results have been also reported at $z\sim1-3$.\cite{urrutia2012,glikman2012,bongiorno2014,zakamska2016} 
To investigate whether \targname\ at $z=7.2$ is also consistent with this scenario, we compare its physical properties with those of dusty starbursts, red quasars, and blue quasars in the literature.  

{In Extended Data Fig.~9}, we compare $L_{\rm [CII]}$ and $L_{\rm IR}$ properties. 
We find that \targname\ has a relatively low $L_{\rm [CII]}$/$L_{\rm IR}$ ratio among all populations, while we confirm that there are two cases whose ratios are comparable in a similar composite system of the AGN and starburst population identified in the local universe. 
We also find that the red quasars at $z\sim3-5$, including super-Eddington accretion red quasar W2246-0526 at $z=4.6$,\cite{tsai2018} have similarly low $L_{\rm [CII]}$/$L_{\rm IR}$\cite{diaz-santos2021}.
These results suggest that the interstellar medium (ISM) conditions of \targname\ are unique compared to similar systems in the local universe and could resemble the super-Eddington accretion red quasars. 
The $L_{\rm [CII]}$/$L_{\rm IR}$ ratio is known to have a strong anti-correlation with $\Sigma L_{\rm IR}$, where the decrease of the \cii\ emissivity is explained by the strong radiation field in the high $\Sigma L_{\rm IR}$ regions\cite{diaz-santos2013, spilker2016,gullberg2018}. 
Based on the upper limit of the effective radius of the 1.3~mm continuum (Section 4), 
we find that \targname\ has a lower limit of $\log(\Sigma L_{\rm IR}) > 13 \,L_{\odot}/{\rm kpc}^{2}$, aligned with the highest-end of the anti-correlation. 
This indicates that the host of \targname\ is one of the most vigorously star-forming systems, which produces the intense radiation field and induces the high $T_{\rm d}$ of 80 $\pm$ 21 K. 

{In Extended Data Fig.~10}, we compare SFR (a), $M_{\rm gas}$ (b), $M_{\rm dust}$ (c), and the depletion time scale $\tau_{\rm depl.}$ (d) between \targname\ and other populations at $z>6$, including the dusty starburst (orange diamond), the blue quasar (blue squares), the red quasar (magenta circle and shaded region), and Lyman-break galaxies (green triangle) from the literature\cite{strandet2017,marrone2018,kim2019,bakx2020,laporte2017,venemans2020,hashimoto2019,hashimoto2019b,kato2020,izumi2021,yang2020,wang2021,novak2019,fan2018,diaz-santos2021}.
For dusty starburst and quasar populations, 
we estimate SFR in the same manner as \targname\ (i.e., conversion from $L_{\rm IR}$). The $L_{\rm IR}$ values are systematically calculated by assuming a single MBB with $T_{\rm d}=47$ K and $\beta_{\rm d}=1.6$ in the same manner as previous studies\cite{kim2019,venemans2020} when the source has been observed only with a single submm or mm band, otherwise taken from the literature\cite{strandet2017,walter2018,hashimoto2019b}.
Because we show \targname\ with $L_{\rm IR,\,SF}$ after subtracting the AGN contribution of 8\%,
we also subtract the same fraction in the $L_{\rm IR}$ estimates for the quasar populations. 
For the dusty starburst at $z=6.9$\cite{strandet2017,marrone2018}, the AGN contribution is estimated to be 0--50\% with nearly equal probability in the literature, where we assume it to be 25\% and evaluate the error with the cases of 0\% and 50\%. 
For the galaxy population, we use SFR estimates compiled in the literature\cite{harikane2020} applying corrections for the different IMF assumptions.
$M_{\rm gas}$ estimates are taken from CO line spectroscopy results in the literature,
while we also derive $M_{\rm gas}$ from $L_{\rm [CII]}$ with $\alpha_{\rm [CII]}= 22 M_{\odot}/L_{\odot}$\cite{zanella2018} for the sources that have not been observed with the CO lines. 
For $M_{\rm dust}$, we systematically calculate by using the equation: 
\begin{equation}
M_{\rm dust} = \frac{F_{\nu} D_{\rm L}^{2}}{(1+z)\kappa_{\nu}(\beta_{\rm d})B_{\nu}(\nu,T_{\rm d})}, 
\end{equation}
where $F_{\nu}$ is the flux density at a frequency of $\nu$, $D_{\rm L}$ is the luminosity distance, $B_{\nu}$ is the Planck function. The dust opacity coefficient is given by $\kappa_{\nu}=5.1(\nu/\nu_{250\mu{\rm m}})^{\beta_{\rm d}}$ cm$^{2}$~g$^{-1}$, which is used in the DL07 model. 
We assume $T_{\rm d}=47$ K and $\beta_{\rm d}=1.6$ when the source has been observed only with a single submm or mm band, otherwise we use the longest wavelength measurements available and adopt the $T_{\rm d}$ and $\beta_{\rm d}$ estimates in the literature\cite{strandet2017,walter2018,hashimoto2019b,bakx2020}. 
Because a $\pm$10 K change in the $T_{\rm d}$ assumption produces a $\sim0.2$--0.3~dex difference in the SFR and $M_{\rm dust}$ estimates for the sources whose $T_{\rm d}$ and $\beta_{\rm d}$ are assumed to be 47~K and 1.6, respectively, we add a 0.2~dex uncertainty to the 1$\sigma$ measurement uncertainty in their error bars in Extended Data Fig.~10.  
We find that the $M_{\rm dust}$ and $M_{\rm gas}$ values of \targname\ fall in the range probed by blue quasars at the same epoch, while the implied SFR is the highest among $z>7$ objects so far observed. 

This indicates that the intense starburst is taking place in the host galaxy of \targname\ with a very short depletion time scale of $\sim$10 Myr, 
which is consistent with the scenario that the red quasar is forming in the dusty starburst.
The higher SFR of \targname's host galaxy than the hosts of lower-redshift red quasars may indicate that \targname\ is experiencing an early stage of its transition phase from the dusty starburst to the blue quasar.
In fact, the $M_{\rm BH}$ values of the lower-$z$ red quasars are estimated to fall in the super-massive regime of $\log(M_{\rm BH}/M_\odot)$ = 9.3--9.6\cite{tsai2018,kato2020}. 
This may suggest that these lower-$z$ red quasars are found at the end phase of the SMBH evolution and that the super-Eddington accretion in W2246-0526 is caused by an active quasar duty cycle even at the end phase. 
Although there is a possibility that the SFR values in the blue quasars are also increased if their $T_{\rm d}$ values are as high as \targname, previous studies of blue quasars at $z>6$ with multi-band FIR photometry in ALMA Bands 6 and 8 generally show the FIR SED with $T_{\rm d}\sim$40--50~K\cite{walter2018,hashimoto2019b,novak2019}. Furthermore, 
the rest-frame IR regime of the blue line in Fig.~2 implies that the AGN contribution to $L_{\rm IR}$ could be much larger in these luminous quasars than that of \targname, which reduces their SFR estimates. 

In Extended Data Fig.~11, we present $M_{\rm BH}$ and $M_{\rm dyn}$ properties. 
For \targname, we show the Eddington-limited $M_{\rm BH}$ estimate in the red circle and the potential $M_{\rm BH}$ range indicated by its extremely faint X-ray property in the red shaded regions whose colour scale and vertical range of each red shade region correspond to those of Fig.~3. 
For blue quasars at $z\sim$6--7, we show $M_{\rm dyn}$ measurements based on a systematic kinematic modeling with the ALMA data\cite{pensabene2020,neeleman2021}. 
We include $M_{\rm dyn}$ measurements for $z\gtrsim6$ quasars based on the assumption of the rotating disc geometry and the axial ratio of \cii\ flux map as proxy of the disc inclination angle\cite{wang2013,willott2015,willott2017,venemans2017,decarli2018,izumi2018,izumi2019}. 
We also show 1) the best-fit relation between the stellar mass in the bulge $M_{\rm bulge}$ and $M_{\rm BH}$ obtained in local quiescent galaxies\cite{kormendy2013}, where we use $M_{\rm bulge}$ as $M_{\rm dyn}$, and 2) the $M_{\rm BH}$ and $M_{\rm star}$ relation for red quasars at $z\sim2$ that generally fall below the local relation\cite{bongiorno2014}. 
We find a relatively low fraction of $M_{\rm\,BH}/M_{\rm\,dyn} <0.2\%$ that falls below the local relation\cite{kormendy2013}, similar to the general relation of the red quasars at $z\sim2$. 
In contrast to ideas of an ``over-massive'' SMBH relative to the host galaxy reported in previous optically-luminous quasars at $z>6$\cite{wang2013,decarli2018,pensabene2020,neeleman2021}, 
the ``under-massive'' SMBH of \targname\ at $z=7.2$ offers an intriguing path to the co-evolution between the SMBH and its host in the early universe: the host galaxies grow earlier than the SMBHs, which is aligned with the predictions of the merger-driven SMBH evolution models\cite{hopkins2008}. 
The ``under-massive'' SMBH is also argued in recent reports of less luminous quasars\cite{willott2015, izumi2018}, 
but generally their central BHs are already massive $M_{\rm\,BH}\sim10^{8-9}\,M_{\odot}$ with low $\lambda_{\rm\,Edd}\sim0.1$--0.2\cite{onoue2019,kim2019}, and thus could be placed at the end-phase of the SMBH evolution after the blue-quasar phase\cite{kim2019}. 
We note that it is unclear whether the quasar host galaxies at $z\gtrsim6$ are the bulge-dominated systems similar to the local quiescent galaxies. We thus also show another best-fit relation between the stellar mass of the entire system and $M_{\rm BH}$ among local AGNs\cite{reines2015}, 
where the $M_{\rm BH}$ range of \targname\ still falls below or on the best-fit relation.

\subsection{9. The IR/Radio correlation}
The correlation between the IR luminosity and the radio emission has been empirically known for several decades. Based on the radio detection of \targname\ at 20~cm/1.5~GHz\cite{owen2018}, we evaluate the IR and radio correlation. 
From previous studies, the IR and radio correlation is typically evaluated with the parameter $q_{\rm IR}$ given by 
\begin{equation}
q_{\rm IR} = \log\left(\frac{1.01\times10^{18}\,L_{\rm IR}}{4\pi D_{L}^{2} [L_{\odot}]}\right) - \log\left(\frac{10^{-32}F_{\rm 1.4GHz}}{(1+z)^{\alpha_{\rm radio}-1}[\mu{\rm Jy}]}\right), 
\end{equation}
where $F_{\rm 1.4GHz}$ is the observed flux
density at 1.4~GHz, and $\alpha_{\rm radio}$ is the radio spectral index which is defined by $F_{\nu}\propto\nu^{\alpha_{\rm radio}}$. 
Due to the non-detection in a deep 10-GHz map\cite{murphy2017}, we obtain a constraint of $\alpha_{\rm radio} < -1.0\pm0.6$. Given the large uncertainty for the $\alpha_{\rm radio}$ constraint, we adopt a typical value of $\alpha_{\rm radio}=-0.75$\cite{ibar2010,murphy2017}, convert the observed 1.5~GHz flux density to the 1.4~GHz flux density, and estimate the $q_{\rm IR}$ value of $2.1\pm0.3$. 
This is consistent with typical value range of local starburst galaxies and high-redshift SMGs\cite{yun2001, magnelli2010} and a recent report of the redshift trend among high-redshift star-forming galaxies\cite{delhaize2017}. 
We thus conclude that the majority of the radio emission of \targname\ is caused by the star-formation and classify the \targname\ as a radio-quiet object. 
This property agrees with the young quasar interpretation of \targname, because the radio loudness is a strong function of $M_{\rm BH}$ and the radio-loud quasars generally have massive $M_{\rm BH}$ $> 10^{9}\, M_{\odot}$\cite{dunlop2003}. 

\subsection{10. Comparison with simulations}

We compare our observational results with predictions from a  data-constrained, semi-analytic model {\tt GAMETE/QSOdust} (GQd) aimed at studying the formation and evolution of high-redshift quasars and their host galaxies in a cosmological framework\cite{valiante2011,valiante2014,valiante2016,ginolfi2019}.
Here we have analysed the hierarchical merger histories of 10 massive dark matter (DM) halos with $M_{\rm halo}=10^{13}\,M_{\odot}$ at $z=6.4$, designed to reproduced the observed properties of the optically-luminous quasar SDSSJ1148 at $z = 6.4$ to investigate if we could identify---among its progenitors at $z=7.2$---systems with physical properties similar to \targname.

We first produce the 10 merger trees for a 10$^{13}$ $M_{\odot}$ dark matter halo, decomposing it into its lower mass progenitors backward in time from $z=6.4$ to $z=24$ using a Monte Carlo algorithm based on the Extended Press-Schechter formalism. 
Then GQd follows the evolution of the baryonic component within each progenitor halo along a merger tree, from $z=24$ down to $z=6.4$. 
At each redshift, in each halo, we follow the formation of stars and BHs (light and heavy seed formation channels are simultaneously implemented in the model) according to the environmental properties (i.e. metallicity of the interstellar medium and the level of illuminating external ionizing and H$_2$ photo-dissociating radiation field). BHs in the centre of galaxies can then grow via gas accretion and mergers with other BHs during major halo-halo mergers (DM halos pair mass ratio $>$ 1:4) while in minor ($<$ 1:4) halo-halo mergers the least massive of the two BHs is ``ejected'', it is considered as a ``satellite'', and we do not further follow its evolution. 
We account for the effect of stellar- and AGN-driven feedback in the form of energy-driven winds 
and include calculations for the IR luminosity from the host galaxy and the X-ray luminosity from accreting BHs, considering the primary component of the hot corona and the reflection component of the surrounding neutral medium.

For each 10$^{13}$ $M_{\odot}$ halo merger tree GQd generates a catalog of progenitor galaxies. 
Each catalog contains the properties (e.g., the mass of BH, gas, stars, dust, SFR, $L_{\rm IR}$, $L_{\rm X}$) of all the progenitor systems (galaxy+BHs).  
In Figure \ref{fig:simulation}a, we show all the progenitors at redshift slices of $z=7.1$, 7.2, and 7.3 in the catalog. 
We mark four progenitors with black circles whose X-ray, optical, and host galaxy properties are close to \targname\ with criteria of $L_{\rm X}<10^{44.5}$ erg~s$^{-1}$, $\alpha_{\rm ox}< -1.9$, and SFR $>$ 100 $M_{\odot}$~yr$^{-1}$.  
The assembly histories of the nuclear BHs for these four progenitors are shown in Figure \ref{fig:simulation}b. 
We find that one of the four progenitors is a direct progenitor which evolves into a luminous quasar harbouring a SMBH with $M_{\rm BH} > 10^{8} M_{\odot}$ at $z=6.4$. 
We also confirm that this progenitor resides in one of the most massive dark matter halos at $z\sim7.2$ with the dark matter halo mass of $\sim 10^{12.2} M_{\odot}$.
In the assembly histories, the remaining three progenitors are subsequently ejected from the centre or become satellites of more massive BHs as a consequence of a minor merger experienced by their host galaxies. Therefore, these BHs are not the direct progenitors of the final SMBHs, although their entire systems also form the SMBHs with $M_{\rm BH}\sim10^{9.5-10} M_{\odot}$ at $z=6.4$. 

\subsection{11. Classical colour selection for high-$z$ quasar search}

A remarkable aspect is that \targname\ is discovered in a relatively small area coverage of the entire \textit{HST} archive ($\sim3$ deg$^{2}$), compared with previous wide-area surveys used in the high-redshift quasar search.  
Mazzucchelli et al. (2017) use Pan-STARRS1\cite{morganson2012} and UKIDSS\cite{lawrence2007} data and select the quasar candidates at $z>6.5$ with the optical colour criteria of\cite{mazzucchelli2017} 
\begin{eqnarray}
z-y &>& 1.4   \\
y-J &\leq& 1.0. 
\end{eqnarray}
Integrating the best-fit SED of \targname\ through the Pan-STARRS1 $z$ and UKIRT $y$ and $J$ filter bandpasses, we find it has $z-y=5.8$ and $y-J=0.5$, which comfortably satisfies the optical--NIR quasar selection criterion. 
This indicates that the quasar population similar to \targname\ could be identified in previous high-$z$ quasar surveys if the data are sufficiently deep, such as The Canada–France High-$z$ Quasar Survey\cite{willott2010} and Subaru High-$z$ Exploration of Low-luminosity Quasars\cite{matsuoka2016,matsuoka2017,matsuoka2018b,matsuoka2019b}. 
\targname\ also meets the optical--NIR colour criteria used in recent discoveries of the luminous quasars at $z\sim7.5$\cite{banados2018,yang2020,wang2021}, 
although the optical--NIR data of the wide-area surveys used in these discoveries is almost 2 orders magnitude shallower than that of the GOODS-North field. 
These colour selection results indicate that the identification of \targname\ at $z=7.2$ in the relatively small area of the entire \textit{HST} archive might be just explained by chance, although the expected probability is less than 1\% from the quasar luminosity function\cite{matsuoka2018} and the red quasar fraction at $z\sim6$.\cite{kato2020}  
There are two other possibilities. 
The first is that the transitioning young quasar more frequently emerges at $z>7$ than at $z=6$. 
The second is that the quasar population similar to \targname\ has been identified in the previous surveys, but not regarded or classified as a quasar due to its faintness in the rest-frame UV, MIR, X-ray, and radio continuum in the follow-up spectroscopy and/or multi-wavelength analyses.  
In fact, the presence of the deep {\it HST} and MIPS data is crucial for the interpretation for \targname\ (Section~5). Without them, the uniquely-faint properties of \targname\ in the rest-frame UV emission lines and X-ray generally conclude its classification as a luminous galaxy.
Recent studies have also suggested, both observationally and theoretically, a potential high abundance
of the dust-rich quasar population at $z > 7$\cite{davies2019, ni2020}. 
A systematic deep, high-resolution optical--MIR imaging campaign for all luminous high-$z$ galaxy candidates could lead to additional discoveries similar to \targname. 
Given the relatively robust calibration\cite{vestergaard2006} and lesser effects from the slim disk, 
detecting broad Balmer emission lines could provide a decisive conclusion for the quasar classification, which will soon become possible even at $z>7$ with the launch of the {\it James Webb Space Telescope}. 
Moreover, even if we do not detect the broad lines with {\it JWST}, the results will suggest further exciting possibilities: 
the existence of an extraordinary UV luminous and compact star-forming region (Extended Data Fig.~\tcb{7}), or that is exactly what the first quasars look like.  

\subsection{Code availability.}
The \textit{HST} and \textit{Spitzer} data were processed with {\sc grizli} and {\sc golfir}, available at \url{https://github.com/gbrammer/grizli} and \url{https://github.com/gbrammer/golfir}, respectively. 
The \textit{HST} F125W image is analyzed with {\sc galfit} which is available at \url{https://users.obs.carnegiescience.edu/peng/work/galfit/galfit.html}. 
The NOEMA data were reduced using the GILDAS software. 
The CASA pipeline version of 5.6 is also used for imaging the NOEMA interferometric data. 
These are available at \url{https://casa.nrao.edu/casa\_obtaining.shtml}. 
\url{https://www.oso.nordic-alma.se/software-tools.php}. 
The online Portable Interactive Multi-Mission Simulator is available at \url{https://heasarc.gsfc.nasa.gov/cgi-bin/Tools/w3pimms/w3pimms.pl}. 

\subsection{Data availability.}

This paper makes use of the following \textit{HST} data from programs 9583, 9727, 9728, 10189, 10339, 11600, 12442, 12443, 12444, 12445, 13063, 13420, 13779, available at  \url{https://archive.stsci.edu/}.  The reduced \textit{HST} and \textit{Spitzer} image mosaics are available at \url{https://doi.org/10.5281/zenodo.4469734}.
Other products from the CHArGE project are avilable at \url{https://gbrammer.github.io/projects/charge/}. 
The NOEMA data that supports our finding consists of  ED19AD and W20EO that are available at \url{https://www.iram-institute.org/EN/content-page-386-7-386-0-0-0.html}. 
The SED of the SDSS quasar at $z=3.11$ used in Fig.~1 is available from the SDSS DR12 website \url{https://dr12.sdss.org/spectrumDetail?plateid=6839&mjd=56425&fiber=146}.  
The SEDs of local quasar and starburst are available from the SWIRE template website \url{http://www.iasf-milano.inaf.it/$\sim$polletta/templates/swire\_templates.html}. 
The datasets generated and/or analyzed during the current study are available from the corresponding author on reasonable request.

\clearpage 
\newpage

\textbf{Extended Data Tables}

\begin{table*}[h]
\label{tab:photometry}
\begin{center}
{\small \textbf{Extended Data Table 1$|$ Multi-wavelength photometry of \targname}}
\begin{tabular}{rccccc}
\hline
Observed $\lambda$&Flux density$^{\ddag}$ &Uncertainty&Telescope&Instrument& reference \\\ 
 [$\mu$m]  & [$\mu$Jy] & [$\mu$Jy]  &        &        &          \\ \hline
0.44    & 0.000      &  0.008 & \textit{HST}  & ACS/F435W  & This work (CHArGE)         \\
0.61    & $-0.006$   &  0.005 & \textit{HST}  & ACS/F606W  & ''         \\
0.78    & 0.005      &  0.007 & \textit{HST}  & ACS/F775W  & ''         \\
0.81    & $-0.011$   &  0.006 & \textit{HST}  & ACS/F814W  & ''         \\
0.85    & 0.056      &  0.010 & \textit{HST}  & ACS/F850LP  & ''         \\
1.05    & 0.683      &   0.036 & \textit{HST}  & WFC3/F105W  & ''         \\
1.25    & 1.307    & 0.067 & \textit{HST}  & WFC3/F125W  & ''         \\
1.40    & 1.783   &  0.092 & \textit{HST}  & WFC3/F140W  & ''         \\
1.60    &  2.103  &  0.107  & \textit{HST}  & WFC3/F160W  & ''         \\
2.15    &  2.778  &  0.044& Subaru  & MORICS/$K_{\rm s}$  & Kajisawa et al. 2011         \\
3.6     &  3.574    & 0.180   & \textit{Spitzer}  & IRAC/ch1  & This work (CHArGE)         \\
4.5     & 3.907    & 0.197 & \textit{Spitzer}  & IRAC/ch2  & ''         \\
5.8     & 4.138    & 0.546 & \textit{Spitzer}  & IRAC/ch3  & ''         \\
8.0     & 4.553  &  0.471  & \textit{Spitzer}  & IRAC/ch4  & ''         \\
24      & 28.1  &  6.6   & \textit{Spitzer}  & MIPS          & Magnelli et al. 2011         \\
100     & $<$ 1050 & 350 & \textit{Herschel}  & PACS          & Liu et al. 2018         \\
160     & $<$ 2850   &  950 & \textit{Herschel}  & PACS          & ''         \\
250     & $<$ 17,100 &  5,700 & \textit{Herschel}  & SPIRE         & Oliver et al. 2012         \\
350     & $<$ 18,300   &  6,100 & \textit{Herschel}  & SPIRE         & ''         \\
500     & $<$ 12,300  &  4,100 & \textit{Herschel}  & SPIRE         & ''         \\
450     & 8,000 & 5,500 & JCMT  & SCUBA2          & This work        \\
850     & 1,800 & 390   & JCMT  & SCUBA2         & Cowie et al. 2017         \\
1,284   & 460  &  94$^\dagger$   &NOEMA  & Band 3         & This work        \\
3,276   & 24.6   &  6.9$^\dagger$ &NOEMA  & Band 1         & This work        \\
30,000  & {0.8} & 1.1  & JVLA  & X Band       &  Murphy et al. 2017   \\
200,000 & 22.4           &  6.4 & JVLA  & L Band         &  Owen 2018 \\ \hline
\end{tabular}
\end{center}
\flushleft{\footnotesize
{$\ddag$
The potential contributions from nearby objects are subtracted, or confirmed to be negligible (Section 5).}\\
$\dagger$ 
The additional uncertainty of the absolute flux calibration is included by $20\%$ and 10$\%$ at 1-mm and 3-mm band, respectively. 
}
\end{table*}

\begin{table*}[h]
\label{tab:phy_prop}
\begin{center}
{\small \textbf{Extended Data Table 2$|$ Measured and derived source properties}}
\begin{tabular}{lll}
\hline
Parameter & Value & Description  \\ \hline
R.A. & 12:36:16.9195 & Right Ascension (J2000) in {\it HST} \\
Decl. & 62:12:32.127 & Declination (J2000) in {\it HST} \\
$z_\mathrm{UV}$ & $7.23\pm0.05$ & Redshift from \textit{HST} grism spectrum \\
$z_{\rm [CII]}$ & $7.1899\pm0.0005$ & Redshift from \cii\ line \\
$\alpha_{\lambda}$    &      {0.1 $\pm$ 0.3}    & Rest-frame UV continuum slope           \\
$L_{\rm 2,500}$ &    ($2.1\pm0.1)\times10^{30}$ erg s$^{-1}$ Hz$^{-1}$        & Monochromatic optical luminosity at rest-frame 2,500 \AA       \\
$L^\prime_{\rm 2,500}$  & ($3.2\pm0.1)\times10^{30}$ erg s$^{-1}$ Hz$^{-1}$          & Dust corrected $L_{\rm 2,500}$ \\ 
$L_{\rm bol}$ & $1.7 \pm 0.1\times10^{46}~\mathrm{erg}~\mathrm{s}^{-1}$  & 
AGN bolometric luminosity (rest-frame 1216\AA--20~cm) \\
$A_{\rm V,\,qso}$  & $0.3\pm0.1$ & Dust attenuation for the quasar component \\
$A_{\rm V,\,host}$ & $>6$      & Dust attenuation for the host galaxy component  \\
$L_{\rm X}$  & $< 3.9\times10^{42}$ erg~s$^{-1}$  & Rest-frame X-ray (2--10 keV) luminosity \\ 
$L_{\rm 2keV}$  & $< 5.1\times10^{24}$ erg~s$^{-1}$ Hz$^{-1}$ & Monochromatic X-ray luminosity at rest-frame 2 keV \\ 
$\alpha_{\rm ox}$  & $< -2.23$ & Optical to X-ray spectral index \\ 
$L_{\rm IR}$ & (1.2 $\pm$ 0.6) $\times\,10^{13}$  $L_{\odot}$  & Rest-frame IR (8--1,000 $\mu$m) luminosity \\
$L_{\rm IR,\,SF}$  & (1.1 $\pm$ 0.5) $\times\,10^{13}$ $L_{\odot}$  & Rest-frame IR luminosity for the host galaxy component  \\
$L_{\rm IR,\,AGN}$ & (1.0 $\pm$ 0.3) $\times\,10^{12}$ $L_{\odot}$  & Rest-frame IR luminosity for the AGN component \\
$f_{\rm AGN}$ & 8 $\pm$ 5 \%   & AGN contribution to $L_{\rm IR}$ \\
$T_{\rm d}$    & 80 $\pm$ 21 K   & Peak dust temperature \\ 
$L_{\rm [CII]}$   & (1.1 $\pm$ 0.3) $\times\,10^{9}$ $L_{\odot}$ &  \cii\ line luminosity     \\
$L_{\rm CO(7-6)}$ & (1.3 $\pm$ 0.7) $\times\,10^{8}$ $L_{\odot}$ &  CO(7-6) line luminosity      \\
FWHM$_{\rm [CII]}$ & $280\pm40$  km~s$^{-1}$ &   FWHM of the \cii\ line       \\
FWHM$_{\rm CO}$    & $770\pm230$  km~s$^{-1}$    &  FWHM of the CO(7-6) line      \\
$L_{\rm CO(6-5)}$ & $<5.0$ $\times\,10^{7}$ $L_{\odot}$   & CO(6-5) line luminosity  (\cii\ line width assumed)  \\
$L_{\rm [CI]2-1}$ & $<7.9$ $\times\,10^{7}$  $L_{\odot}$  & \ci\ line luminosity  (\cii\ line width assumed)  \\
SFR             & $1,600\pm700$ $M_{\odot}$ yr$^{-1}$   & SFR of the host galaxy \\  
$M_{\rm dust}$        &  (1.6 $\pm$ 1.1) $\times\,10^{8}$ $M_{\odot}$ & Dust mass of the host galaxy \\ 
$M_{\rm gas}$  &  (2.0 $\pm$ 1.2) $\times\,10^{10}$ $M_{\odot}$ & Gas mass of the host galaxy \\ 
$M_{\rm dyn}$  &  (4.5 $\pm$ 0.9) $\times\,10^{10}$ $M_{\odot}$ & Dynamical mass \\
$M_{\rm star}$   & (2.5 $\pm$ 1.4) $\times\,10^{10}$ $M_{\odot}$  & Stellar mass of the host galaxy from $M_{\rm dyn}- M_{\rm gas}$   \\
$q_{\rm IR}$          &   2.1 $\pm$ 0.3            & IR/radio correlation \\
$\alpha_{\rm radio}$ & $< -1.0\pm0.6$ & Radio spectral index   \\ 
$r_{\rm e, FIR}$   & $<$ {0.48} kpc      & Effective radius of the rest-frame FIR continuum \\
$r_{\rm e, [CII]}$  & 1.4 $\pm$ 0.2 kpc & Effective radius of the \cii\ line   \\
\hline
\end{tabular}
\end{center}
\vspace{-0.3cm}
 \vspace{-0.5cm}
\end{table*}

\clearpage 
\newpage

\textbf{Extended Data Figures}

\begin{figure*}[h]
\begin{center}
\includegraphics[angle=0,width=1.0\textwidth]{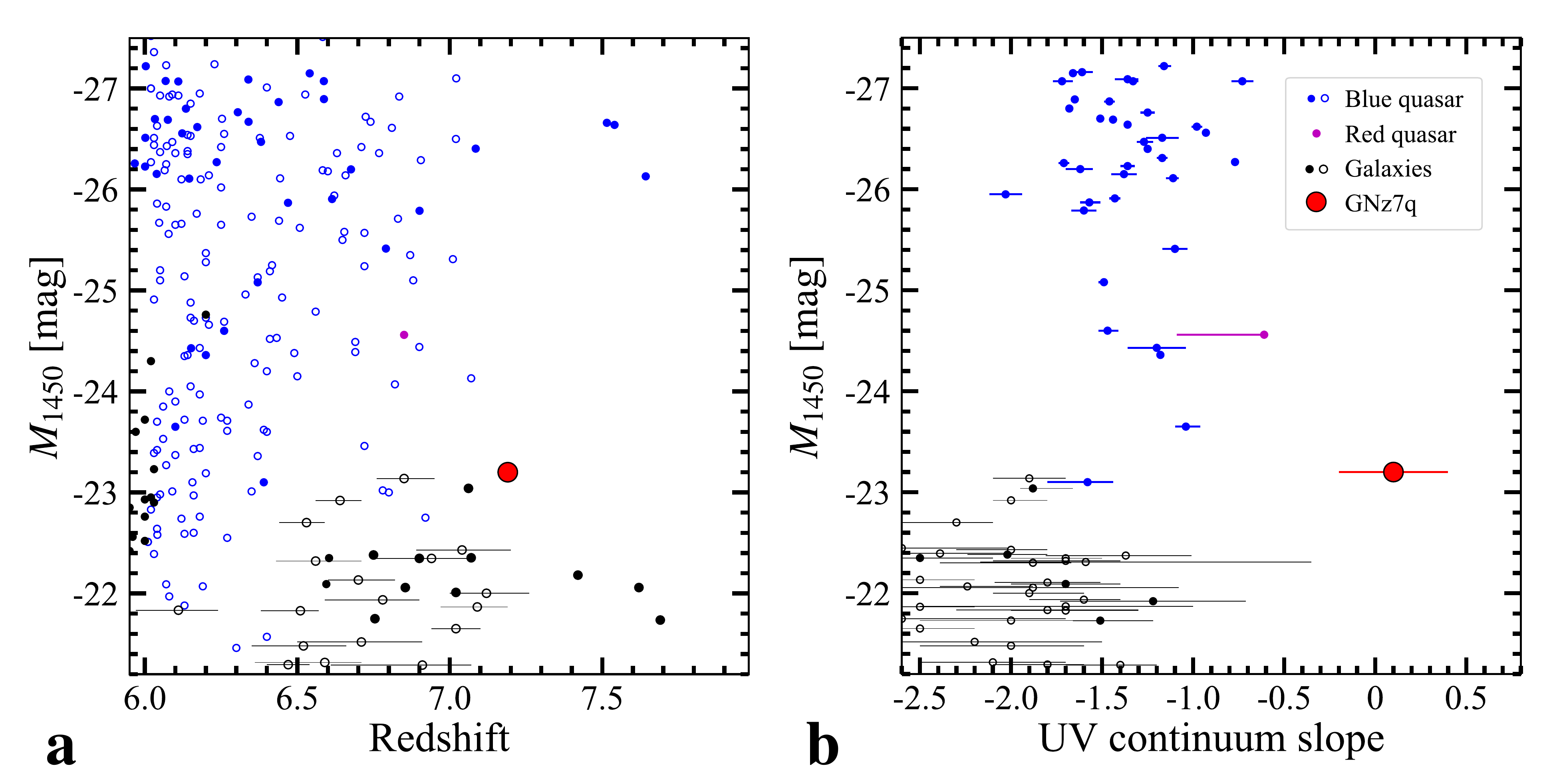}
\end{center}
\small \textbf{Extended Data Figure 1 $|$ Rest-frame UV properties of \targname}. 
The rest-frame 1450 ${\rm \AA}$ luminosity as a function of redshift ({\bf a}) and the UV continuum slope ({\bf b}). 
\targname\ falls between the typical luminosity ranges of quasars and galaxies in the literature\cite{shindler2020,inayoshi2020,yang2020,wang2021}, where both faint quasars and luminous galaxies have been also identified\cite{willott2015,matsuoka2016,matsuoka2017,matsuoka2018b,matsuoka2019,matsuoka2019b,bowler2017,schouws2021}. 
\targname\ shows the reddest UV continuum slope among both galaxies and quasars at $z>6$. 
The galaxies without spectroscopic redshifts and the quasars without a UV continuum slope measurement are displayed in the open symbols. 
The error bars denote the 1$\sigma$ measurement uncertainty. 
\label{fig:muv-color}
\end{figure*}

\begin{figure*}[h]
\begin{center}
\includegraphics[angle=0,width=0.85\textwidth]{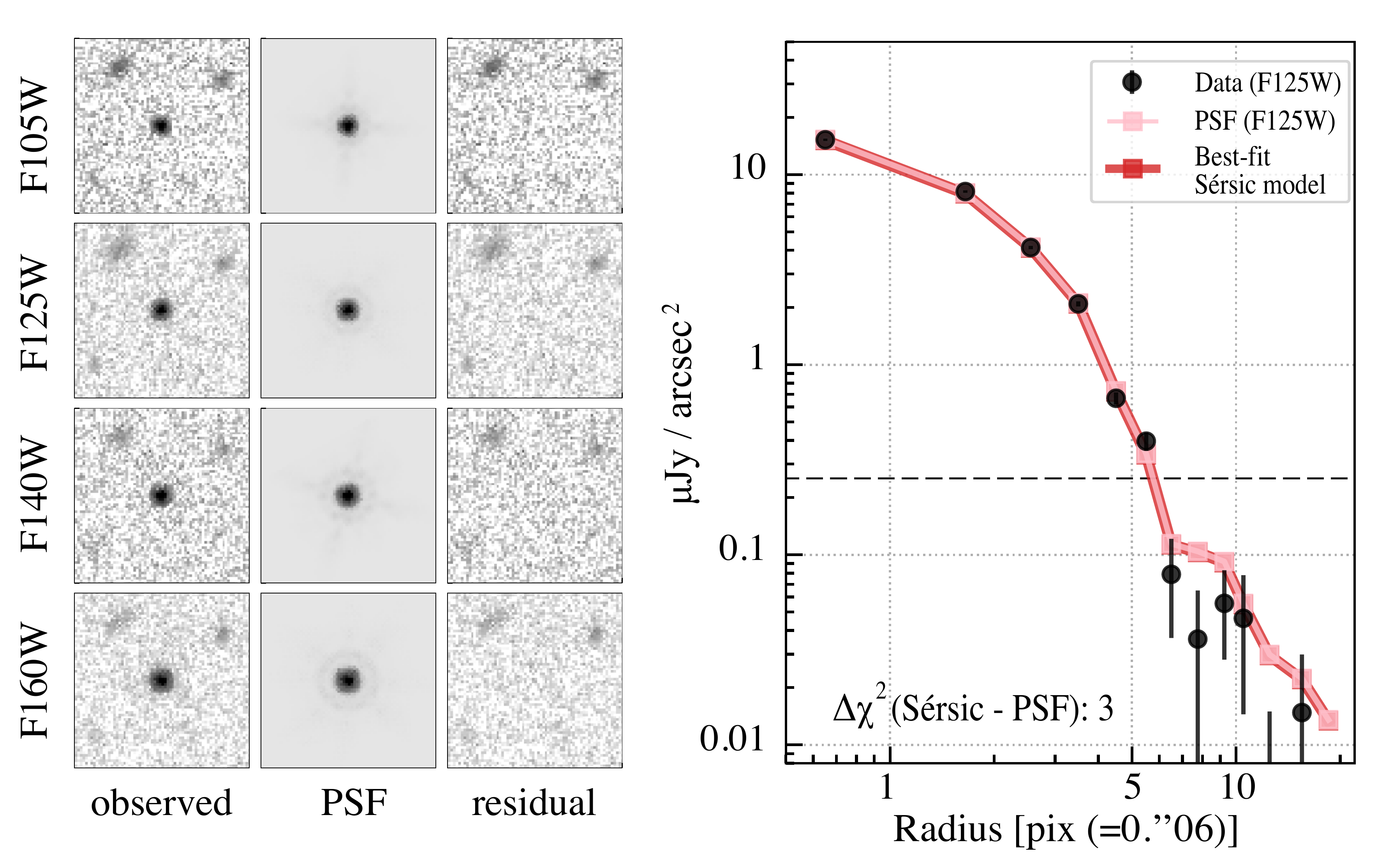}
\end{center}
{
\small \textbf{Extended Data Figure~2 $|$ Point-source morphology of \targname}. 
{\bf a}, 
 HST $4''\times4''$ cutout in the HST WFC3/IR filters of F105W, F125W, F140W, and F160W (left), 
instrumental point spread function (PSF) models\cite{anderson2016} (centre), 
and PSF fit residuals (right). 
{\bf b}, Radial profile for the rest-frame UV continuum of \targname\ observed in F125W.
The black circles show the observed values, while the dark and light red squares and lines present the PSF and the best-fit S\'ersic models (see Methods). 
The error bars denote the 68th percentile in each annulus, 
and the dotted line indicates the standard deviation of the pixel. 
} 
\label{fig:psf}
\end{figure*}

\begin{figure*}[h]
\begin{center}
\includegraphics[angle=0,width=1.0\textwidth]{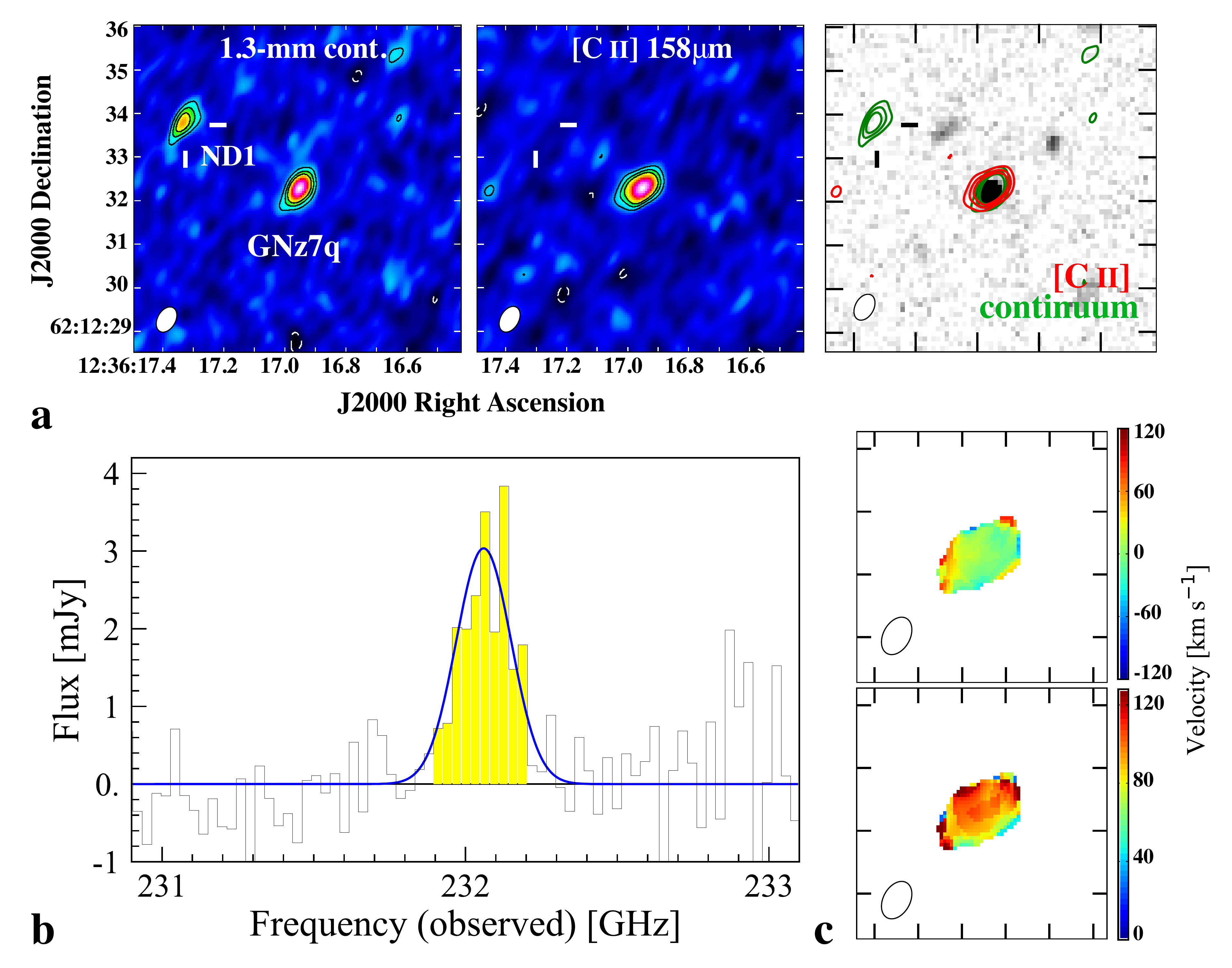}
\end{center}
\small \textbf{Extended Data Figure~\tcb{3} $|$ NOEMA 1-mm observation results}.  
\textit{\bf a,} 1.3-mm continuum (left) and the velocity-integrated \cii\ maps (middle) with the natural weighting. 
We identify a nearby continuum object with a $\sim3''$ offset from \targname\ at the northern east part, dubbed ``ND1''. 
The intensity of the 1.3-mm continuum and the velocity integrated \cii\ is shown in the right panel in green and red contours, respectively, overlaid on the HST/F160W $4''\times4''$ cutout. 
The solid contours are drawn at 3$\sigma$, 5$\sigma$, and 7$\sigma$ levels, 
while the dashed white contours are drown at $-3\sigma$ level. 
The NOEMA synthesized beam is presented at the left bottom. 
\textit{\bf b,} \cii\ line spectrum within a $1.''0$ radius aperture. 
The blue curve is the best-fit Gaussian for the \cii\ line. 
The yellow shaded indicates the velocity range of [$-200:+200$] km~s$^{-1}$ used for the velocity-integrated map in panel {\bf a}. 
\textit{\bf c,} \cii\ line kinematics. 
The top and bottom panel present the velocity-weighted and the velocity-dispersion maps ($4''\times4''$), respectively. 
\label{fig:noema_1mm}
\end{figure*}

\begin{figure*}[h]
\begin{center}
\includegraphics[angle=0,width=1.0\textwidth]{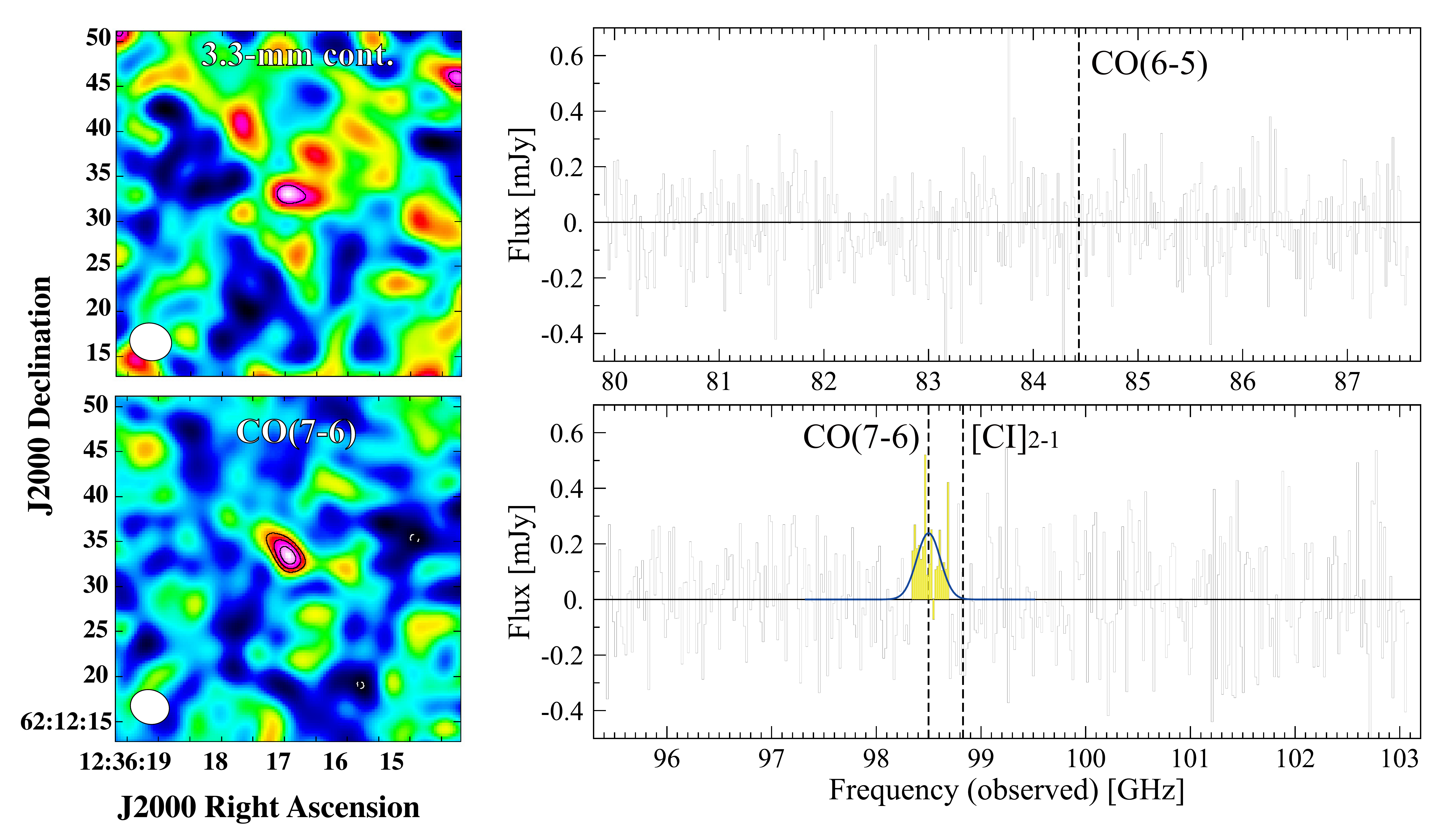}
\end{center}
\small \textbf{Extended Data Figure~\tcb{4} $|$ NOEMA 3-mm observation results}.  
\textit{\bf Left,} 3.3-mm continuum (top) and the velocity-integrated CO(7-6) maps with the natural weighting. 
The black (white) contours are drown at 3$\sigma$, 4$\sigma$, and 5$\sigma$ ($-3\sigma$). 
 \textit{\bf Right,} NOEMA 3-mm band spectrum for LSB (top) and USB (bottom) with a $2.''0$ radius aperture. 
The dashed vertical line indicates the observed frequency of the expected far-IR lines based on the source redshift of $z=7.1899$ determined by the \cii\ line. 
The blue curve is the best-fit Gaussian for the CO(7-6) line. 
The yellow shade indicates the velocity range used for the velocity-integrated map in the left panel. 
\label{fig:co-cont}
\end{figure*}

\begin{figure*}[h]
\begin{center}
\includegraphics[angle=0,width=1.0\textwidth]{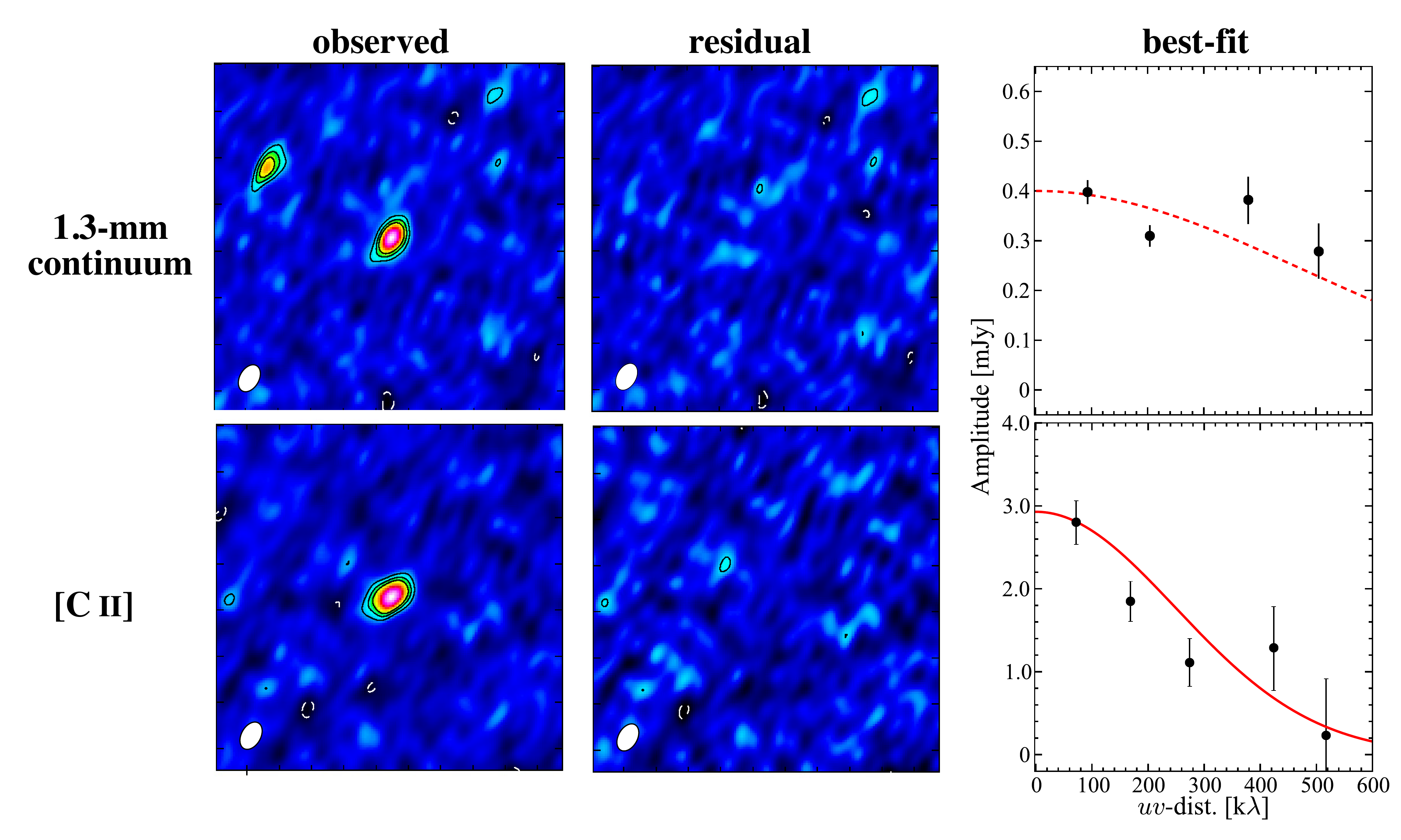}
\end{center}
\small \textbf{Extended Data Figure~\tcb{5} $|$ 1.3-mm continuum (top) and \cii\  (bottom) size measurement results}.  
\textit{\bf Left,} Observed map, which is the same as Extended Fig.~3a. 
\textit{\bf Middle,} Residual map by subtracting the best-fit model visibility obtained with {\sc uvmodelfit}.
{For the dust continuum, we subtract the best-fit model visibility by fixing the major-axis effective radius as the upper limit value of $r_{\rm e, FIR}=0.48$ kpc. 
The visibility of ND1 is subtracted by assuming its profile as a point source before running {\sc uvmodelfit}.
}
\textit{\bf Right,} Amplitude as a function of $uv$ distance. 
The black circles shows the observed visibility. 
The error bars show the standard error of the mean in each $uv$ distance bin. 
The red curve denotes the best-fit $uv$ model for the \cii\ line, while the red dashed curve for the dust continuum indicates the $uv$ model with the upper limit size.
\label{fig:cii-cont}
\end{figure*}

\begin{figure*}[h]
\begin{center}
\includegraphics[angle=0,width=0.72\textwidth]{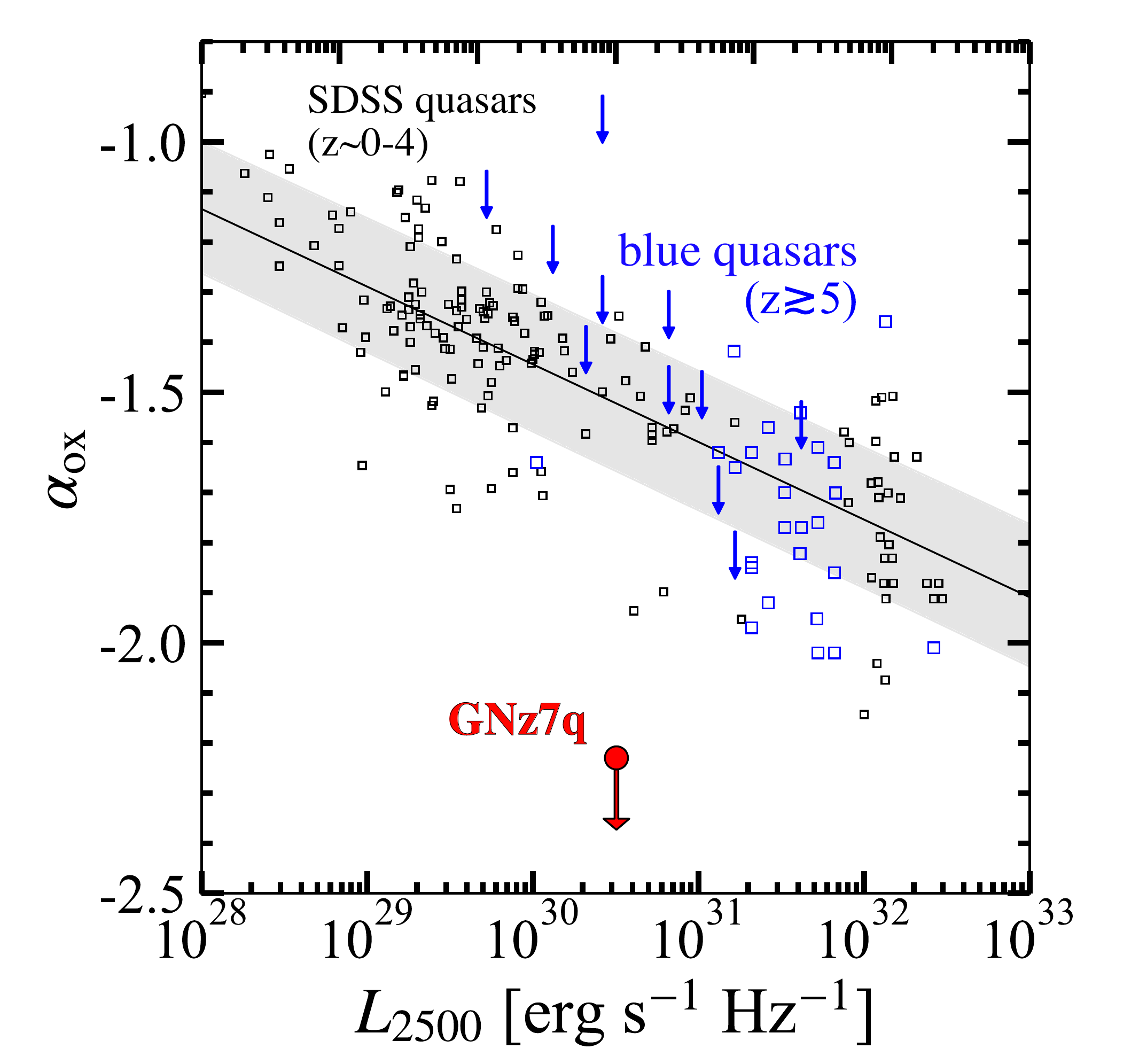}
\end{center}
\small \textbf{Extended Data Figure~\tcb{6} $|$ Optical luminosity vs. $\alpha_{\rm ox}$ correlation}. 
The black and blue squares denote SDSS quasars\cite{just2007,lusso2010,lusso2016} at $z\sim$0--4 and blue quasars\cite{shemmer2006,nanni2017,vito2019} at $z>5$ respectively, taken from the literature. 
The arrows present the upper limits.
The black line represents the best-fit relation based on 1544 quasars taken from the literature\cite{nanni2017}. 
The gray shaded region denotes the 68th percentile derivation, evaluated by propagating the 1$\sigma$ uncertainties of the parameters that define the best-fit relation.
The $\alpha_{\rm ox}$ upper limit of \targname\ (99\% confidence level) is estimated after the extinction correction and deviated from the best-fit relation by more than $5\sigma$.
\end{figure*}

\begin{figure*}[h]
\begin{center}
\includegraphics[angle=0,width=0.75\textwidth]{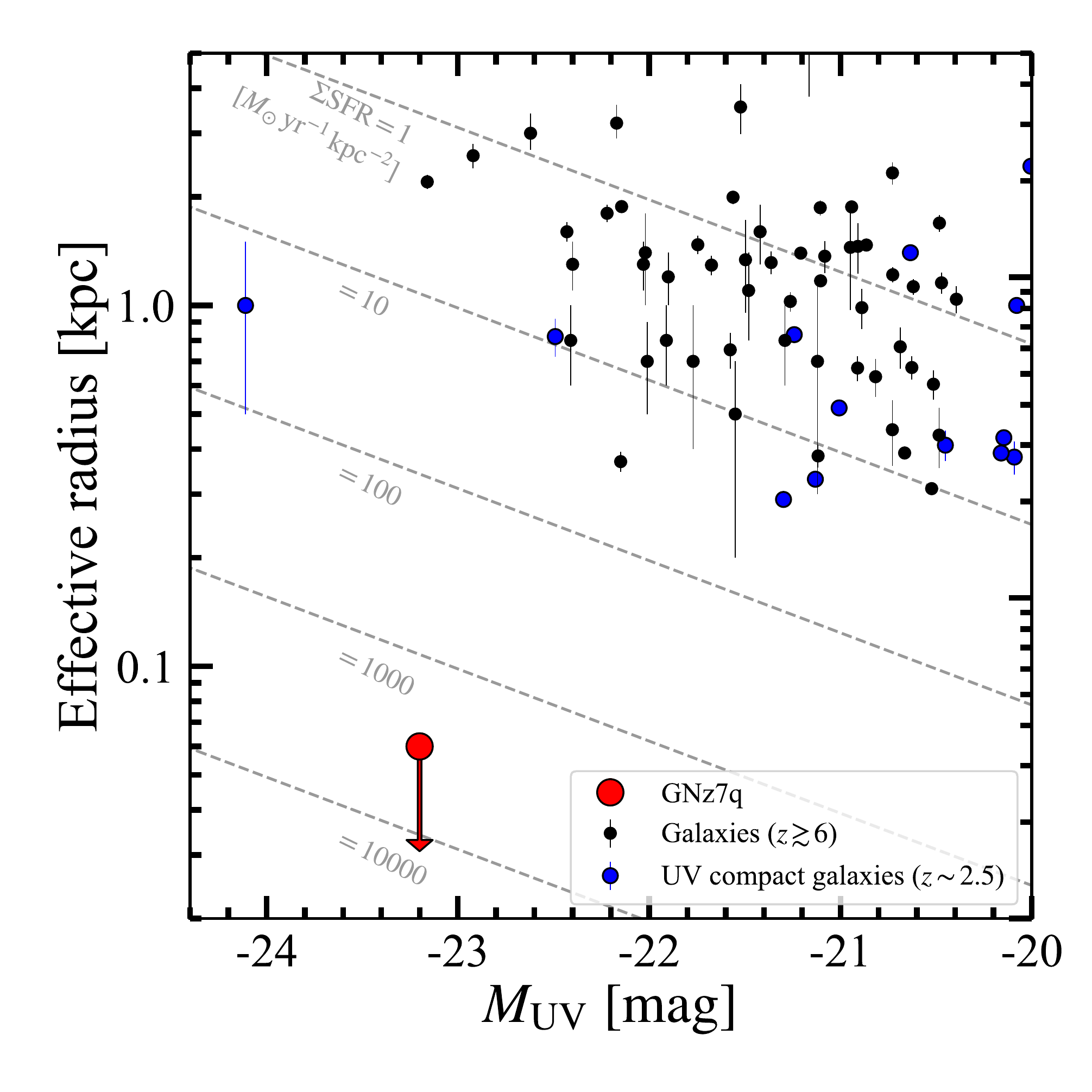}
\end{center}
\small \textbf{Extended Data Figure~\tcb{7} $|$ Rest-frame UV size and luminosity relation}. 
The black and blue circles show the rest-frame UV size measurements in the literature for galaxies\cite{shibuya2015,bowler2017} at $z>5.5$ and for compact galaxies reported at $z\sim2$--3,  respectively\cite{barro2014,marques2020}, but no objects similarly compact and luminous to \targname\ have been identified. 
The error bar denotes the 1$\sigma$ measurement uncertainty, and the sources whose errors exceed the measurements are not presented. 
The dashed line indicates the SFR surface density ($\Sigma_{\rm SFR}$) by converting the UV luminosity to SFR\cite{murphy2011}.
If the compact UV emission in \targname\ is attributed to the star-forming activity, $\Sigma_{\rm SFR}$ reaches $\gtrsim$ 5,000 $M_{\odot}$~yr$^{-1}$~kpc$^{-2}$. 
Note that the UV luminosity is the observed value, and thus $\Sigma_{\rm SFR}$ of \targname\ after dust correction will be more extreme in the star-forming scenario.  
\label{fig:muv-size}
\end{figure*}

\begin{figure*}[h]
\begin{center}
\includegraphics[angle=0,width=1.0\textwidth]{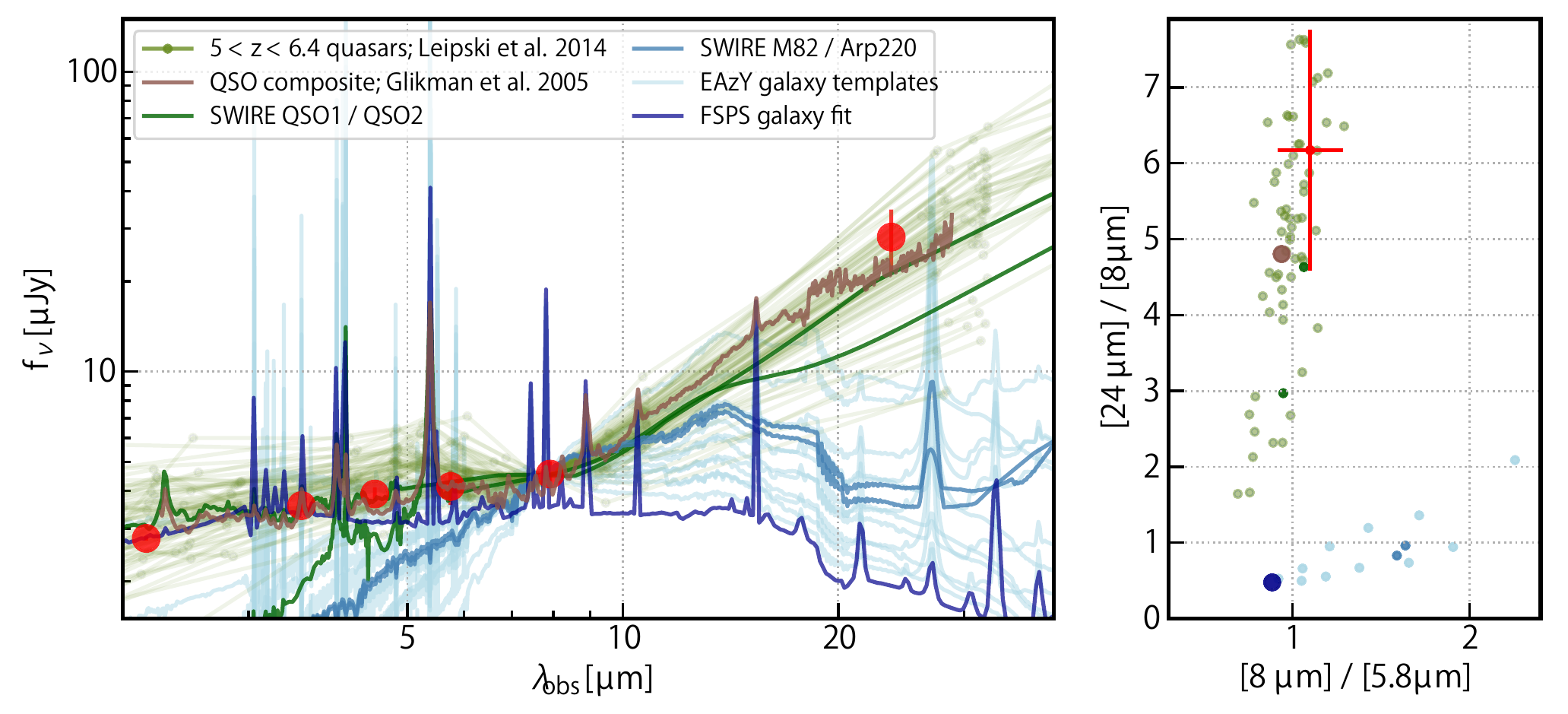}
\end{center}
\small \textbf{Extended Data Figure~\tcb{8} $|$ NIR--MIR SED of \targname}. 
\textit{Left:} Observed-frame SED of \targname\ traced by the \textit{Spitzer} IRAC and MIPS 24~\mum\, bands. The dark blue curve is the best-fit galaxy template (stellar continuum plus nebular emission from ionized gas in H\textsc{ii} regions) constrained at $\lambda_\mathrm{obs} < 10\,\mu\mathrm{m}$.  The thin light blue curves are additional galaxy templates that largely span the galaxy color space at lower redshifts\cite{brammer2008}, and the thicker light blue curves are templates of nearby dusty starbursts M82 and Arp220\cite{polletta2007}.  
The thick green curves are templates of Type 1 and 2 quasars\cite{polletta2007}, and the brown curve is a composite spectrum of nearby quasars\cite{glikman2006}.
The light green curves show the broad-band SEDs of high-redshift quasars at $5<z<6.4$\cite{leipski2014} interpolated to the redshift of \targname. 
Other than the galaxy fit, all SEDs and templates are normalized to the observed 8~\mum\, flux density of \targname.   
\textit{Right:} 
Observed-frame MIR flux ratio diagram for the flux densities at 5.8~\mum, 8~\mum, and 24~\mum\, as observed for \targname\ and integrated from the SEDs displayed in the left panel. 
No templates from stars and star formation alone (blue curves and points) can reproduce the flux enhancement at 24~\mum\, (rest-frame 3~\mum) of \targname, which is fully consistent with the colors of luminous quasars at both low and high redshifts and likely arises from hot dust associated with an active nucleus.
The error bars are obtained by propagating the 1$\sigma$ measurement uncertainty of each photometry. 
\label{fig:mir-colour}
\end{figure*}

\begin{figure*}[h]
\begin{center}
\includegraphics[angle=0,width=1\textwidth]{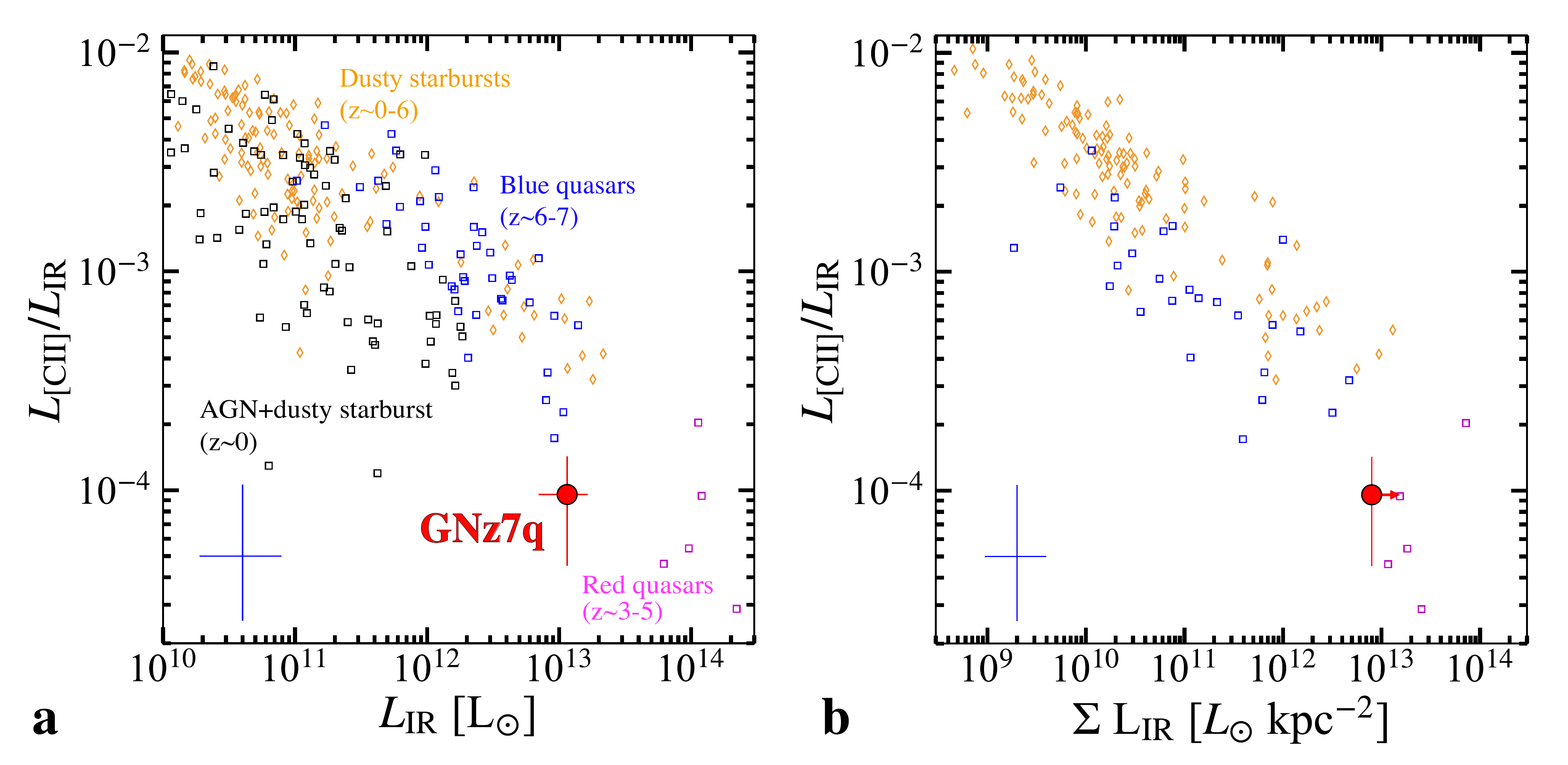}
\end{center}
\small \textbf{Extended Data Figure~9 $|$ 
$L_{\rm [CII]}$ and $L_{\rm IR}$ properties compared with other populations.  
}
We show $L_{\rm [CII]}/L_{\rm IR}$ as a function of $L_{\rm IR}$ ({\bf a}) and $\Sigma L_{\rm IR}$ ({\bf b}). 
For comparison, we also show observational results of local composite systems of AGN and starburst (black square), dusty starbursts at $z\sim$0--7 (orange diamond), blue quasars at $z\sim$6--7 (blue square), and red quasars at $z\sim$3--5 (magenta square) taken from the literature\cite{diaz-santos2013, spilker2016, riechers2013, sargsyan2012, decarli2018, novak2019, yang2020, wang2021, diaz-santos2021}. 
\targname\ is at the extreme end of the relationship painted by known starbursts and quasars.
The $L_{\rm IR}$ values of the blue quasars are calculated by assuming the single modified blackbody ($T_{\rm d}=$ 47 K; $\beta_{\rm d}=$ 1.6), {where the blue bar at the bottom left of the left panel shows a potential error scale with a change of $T_{\rm d}$ by $\pm10$ K from the assumption. 
For \targname, the error bar is obtained by propagating the 1$\sigma$ uncertainties of $L_{\rm [CII]}$ and $L_{\rm IR}$}. 
\label{fig:cii_lir}
\end{figure*}

\begin{figure*}[h]
\begin{center}
\includegraphics[angle=0,width=1.\textwidth]{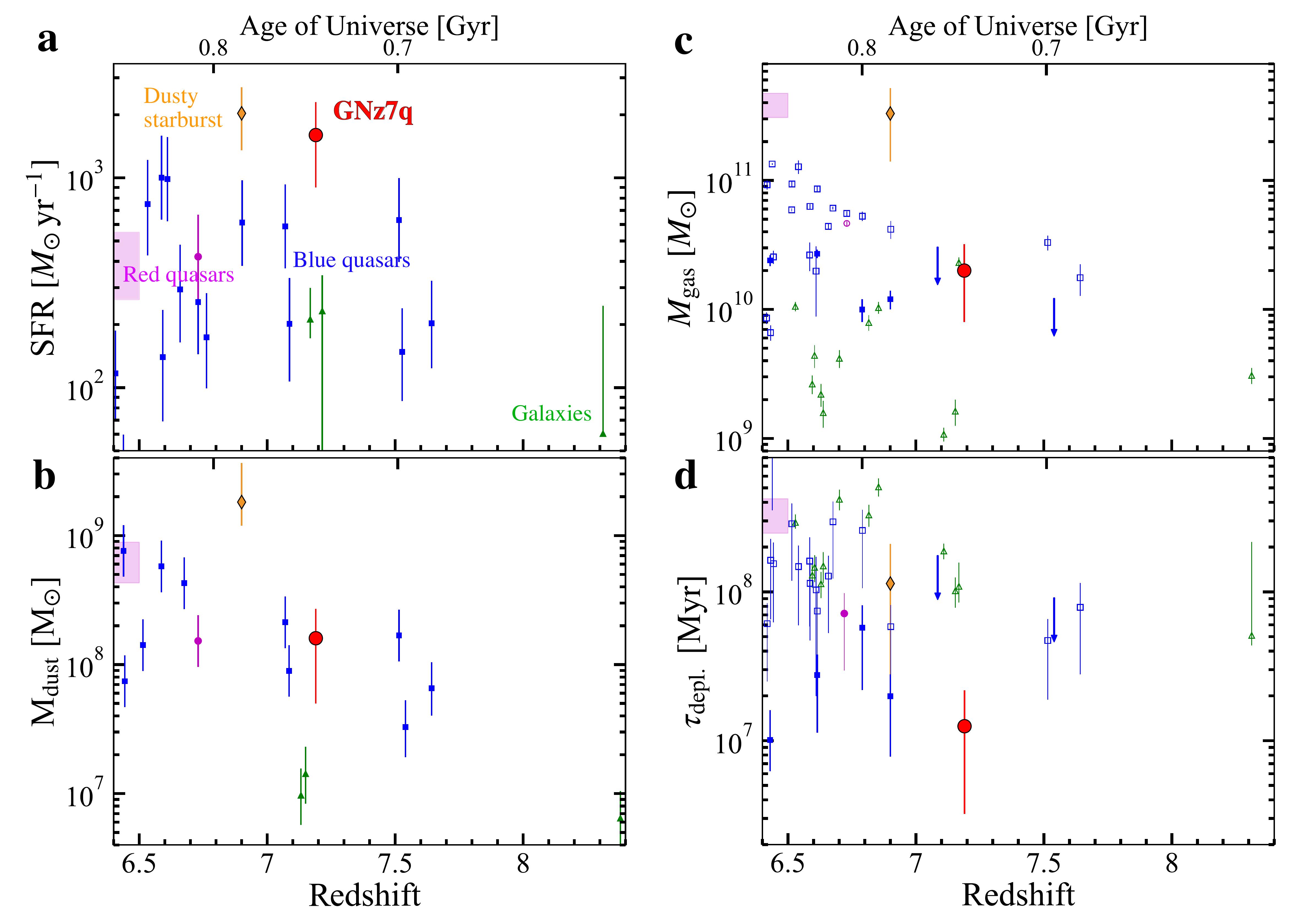}
\end{center}
\vspace{-0.2cm}
\small 
\textbf{Extended Data Figure~10 $|$ Host galaxy properties compared with other populations at $z>6$}.  
{\bf a--d,} 
We show (a) SFR, (b) $M_{\rm dust}$, (c) $M_{\rm gas}$, (d) and $\tau_{\rm depl.}$ as a function of redshift. 
For comparison, we also show other galaxy populations with spectroscopic redshifts: blue quasars (blue square), 
red quasars (magenta circle and shaded region), 
Lyman-break galaxies (green triangle), 
and a dusty starburst galaxy (orange circle) that are taken from the literature\cite{strandet2017,marrone2018,kim2019,bakx2020,laporte2017,venemans2020,hashimoto2019,hashimoto2019b,kato2020,izumi2021,yang2020,wang2021,novak2019,fan2018,diaz-santos2021}.
The magenta shade represents the 68th percentile of the host galaxy properties of the super-Eddington accretion red quasar, W2246-0526, at $z=4.6$\cite{fan2018,diaz-santos2021}. 
The host galaxy of \targname\ show the most vigorously star-forming system at $z>7$ with the large gas reservoir.  
The filled and open symbols in panel {(c)} denote $M_{\rm gas}$ estimates from CO and \cii\ lines, respectively. 
The error bars of SFR and $M_{\rm dust}$ are estimated by propagating the 1$\sigma$ measurement uncertainty and a 0.2-dex uncertainty of the $T_{\rm d}$ assumption, when they are derived from a single submm-mm band (Section 8). 
The error bars of $M_{\rm gas}$ and $\tau_{\rm depl.}$ are estimated with the 1$\sigma$ measurement uncertainty and the propagation from both SFR and $M_{\rm gas}$ uncertainties, respectively.  
For all populations, the different assumptions of the initial mass function and the dust opacity coefficient among the literature are corrected.
\label{fig:host_properties}
\end{figure*}
\newpage

\begin{figure*}[h]
\begin{center}
\includegraphics[angle=0,width=0.75\textwidth]{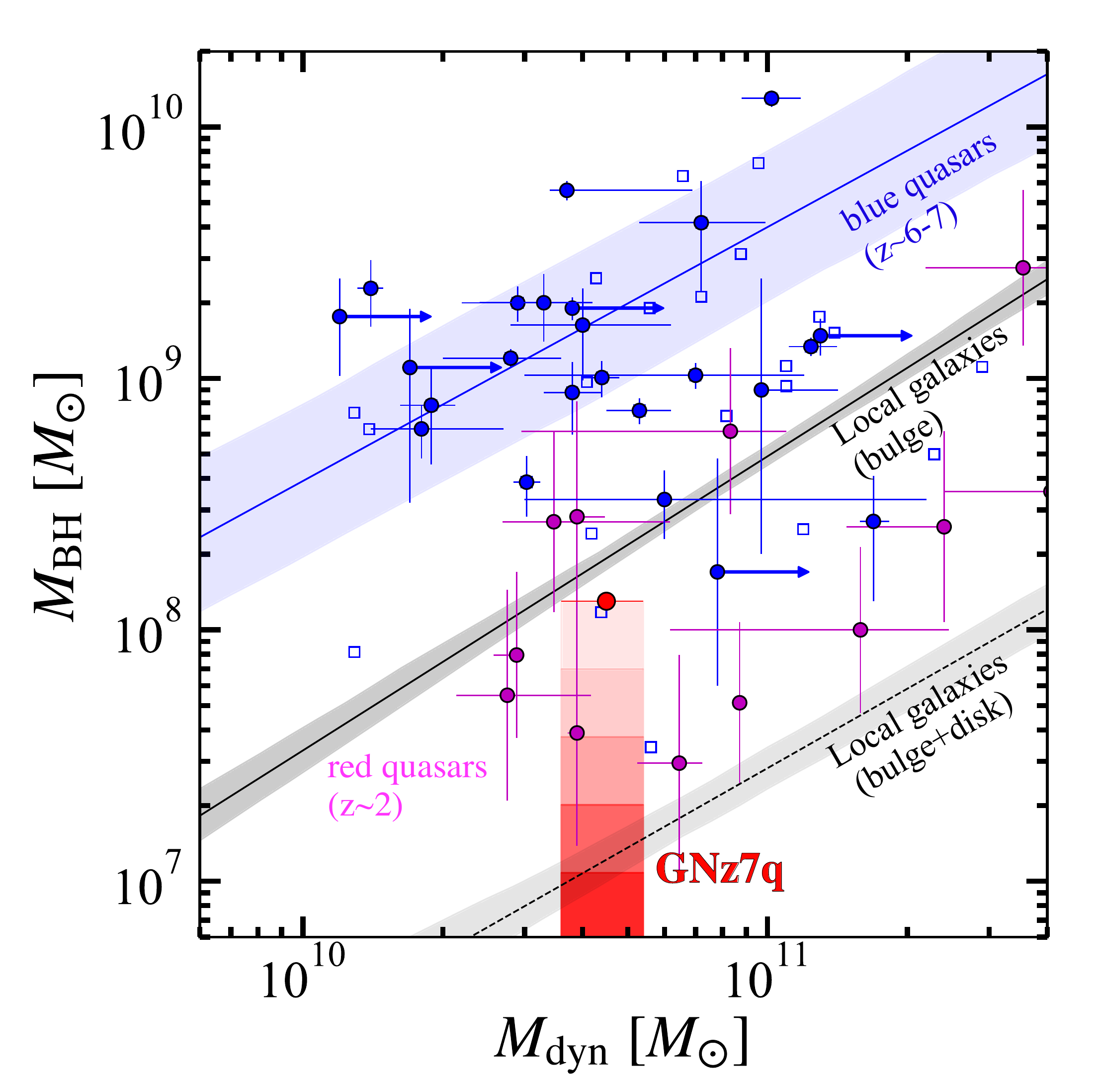}
\end{center}
\small \textbf{Extended Data Figure~11 $|$ $M_{\rm dyn}$ and $M_{\rm BH}$ relation}.  
The colour scale and the vertical range of red-shade regions correspond to those of Fig.~3. 
The red circle and the red-shade regions show the potential $M_{\rm BH}$ range of \targname\ suggested by its faint $L_{\rm bol}$ and extremely faint X-ray property, respectively. 
The horizontal range of the red-shade regions indicates the 68th percentile of the $M_{\rm dyn}$ estimate from the \cii\ line. 
For comparison, we also present $M_{\rm BH}$ and $M_{\rm dyn}$ (or $M_{\rm star}$) estimates for blue quasars at $z\sim6$--7 (blue squares)\cite{willott2015,izumi2019,wang2013,venemans2017,decarli2018,pensabene2020,shindler2020,neeleman2021} and red quasars at $z\sim2$ (magenta circles)\cite{bongiorno2014}.
The error bars denote the 1$\sigma$ uncertainties taken from the literature.
The $M_{\rm dyn}$ values from the kinematic analysis based on the 3D modeling are shown in the filled blue squares with the 1$\sigma$ error bars\cite{pensabene2020,neeleman2021}. 
The $M_{\rm dyn}$ measurements based on the rotation-disk assumption in the literature are shown by the open blue squares. 
The best-fit relation 
for the filled blue squares is shown by the blue line\cite{pensabene2020}.
The black solid line represents the best-fit relation between the bulge mass and $M_{\rm BH}$ among local quiescent galaxies\cite{kormendy2013}. 
The black dashed line denotes the best-fit relation between the stellar mass of the entire system and $M_{\rm BH}$ among local AGNs\cite{reines2015}. 
The shaded regions present the 1$\sigma$ confidence level for the best-fit relations. 
\label{fig:mdyn-mbh}
\end{figure*}
\end{methods}
\end{document}

%% file: _acknowledgement.tex
\subsection{Acknowledgements}

We thank 
M. Onoue, K. Ichikawa, Y. Harikane, and Y. Ono for discussions on the physical properties of \targname\ and the AGN fraction among the brightest Lyman-break galaxies at $z\sim7$; 
E. Murphy and F. Owen for sharing their VLA data; 
D. Marrone for sharing the best-fit SED model of SPT0311-58W; 
K. Whitaker for a helpful advise on writing the manuscript. 
This work is based on the archival data of {\it Hubble Space Telescope}, {\it Spitzer}, {\it Chandra}, Subaru, {\it Herschel}, James Clerk Maxwell Telescope, and The Karl G. Jansky Very Large Array, and the observations of IRAM/NOEMA interferometer (program ID: E19AD and W20EO). 
We acknowledge support from: the Danish National Research Foundation under grant No. 140; the European Research Council (ERC) Consolidator Grant funding scheme (project ConTExt, grant No. 648179); Independent Research Fund Denmark grants DFF--7014-00017; DFF-8021-00130; the Villum Fonden research grant 37440, ``The Hidden Cosmos''\tcr{; the European Union’s Horizon 2020 research and innovation program under the Marie Sklodowska-Curie grant agreement No. 847523 ``INTERACTIONS''}.

%% file: _author_contribution.tex
\subsection{Author contributions}

G.B. reduced and analyzed the optical--NIR data of \textit{HST} and \textit{Spitzer} and discovered \targname.  
S.F., G.B., S.T., G.M., D.W., F.V., C.S., J.F., L.C., R.M. M.V., and F.W. discussed and planned the follow-up observing strategy and the data analysis. 
G.B., G.M., and V.K. conducted the SED analysis and wrote the relevant Methods section. G.B. produced Figs. 1, 2, and Extended Data Figs. 2 and 8. 
D.W. analyzed the X-ray properties from the {\it Chandra} data and wrote the relevant Methods section, 
T.G. reduced and analyzed the SCUBA2 data, and M.K. and I.C. reduced the NOEMA data. 
R.V., M.G., and R.S. performed the cosmological semi-analytical simulation \texttt{GAMETE/QSOdust} 
and wrote the relevant Methods section. 
F.R. worked on the 3D modeling for the NOEMA \cii-line data cube. 
P.O. investigate the properties of the dust-continuum object identified near \targname. 
All authors discussed the results and commented on the manuscript. 
S.F. led the team, being Principal Investigator of the follow-up NOEMA programs, analyzed the NOEMA data, wrote the main text and the Methods section, produced Figs. 3 and 4, Extended Data Tables, and Figs. 1, 3--7, 9--11.